\documentclass[letterpaper,twocolumn,10pt]{article}
\usepackage{usenix-2020-09}
\usepackage{cite}
\usepackage{amsmath,amssymb,amsfonts}
\usepackage{algorithmic}
\usepackage{graphicx}
\usepackage{textcomp}
\usepackage{xcolor}
\usepackage{booktabs}
\usepackage{caption}
\usepackage{subcaption}
\def\BibTeX{{\rm B\kern-.05em{\sc i\kern-.025em b}\kern-.08em
    T\kern-.1667em\lower.7ex\hbox{E}\kern-.125emX}}

\makeatletter
\newcommand{\linebreakand}{%
  \end{@IEEEauthorhalign}
  \hfill\mbox{}\par
  \mbox{}\hfill\begin{@IEEEauthorhalign}
}
\makeatother

\usepackage{cleveref}

\begin{document}

\title{Autonomous Attack Mitigation for Industrial Control Systems
}

\author{
{\rm John Mern}\\
Stanford University
\and
{\rm Kyle Hatch}\\
Stanford University
\and
{\rm Ryan Silva}\\
Johns Hopkins University Applied Physics Lab
\and
{\rm Cameron Hickert}\\
Johns Hopkins University Applied Physics Lab
\and
{\rm Tamim Sookoor}\\
Johns Hopkins University Applied Physics Lab
\and
{\rm Mykel J. Kochenderfer}\\
Stanford University
} 


\maketitle

\begin{abstract} 
Defending computer networks from cyber attack requires timely responses to alerts and threat intelligence.
Decisions about how to respond involve coordinating actions across multiple nodes based on imperfect indicators of compromise while minimizing disruptions to network operations. 
Currently, playbooks are used to automate portions of a response process, but often leave complex decision-making to a human analyst. 
In this work, we present a deep reinforcement learning approach to autonomous response and recovery in large industrial control networks. 
We propose an attention-based neural architecture that is flexible to the size of the network under protection. 
To train and evaluate the autonomous defender agent, we present an industrial control network simulation environment suitable for reinforcement learning.
Experiments show that the learned agent can effectively mitigate advanced attacks that progress with few observable signals over several months before execution. 
The proposed deep reinforcement learning approach outperforms a fully automated playbook method in simulation, taking less disruptive actions while also defending more nodes on the network.
The learned policy is also more robust to changes in attacker behavior than playbook approaches. 
\end{abstract}


\section{Introduction} 

\begin{figure*}[t]
    \centering
    \includegraphics[width=0.75\textwidth]{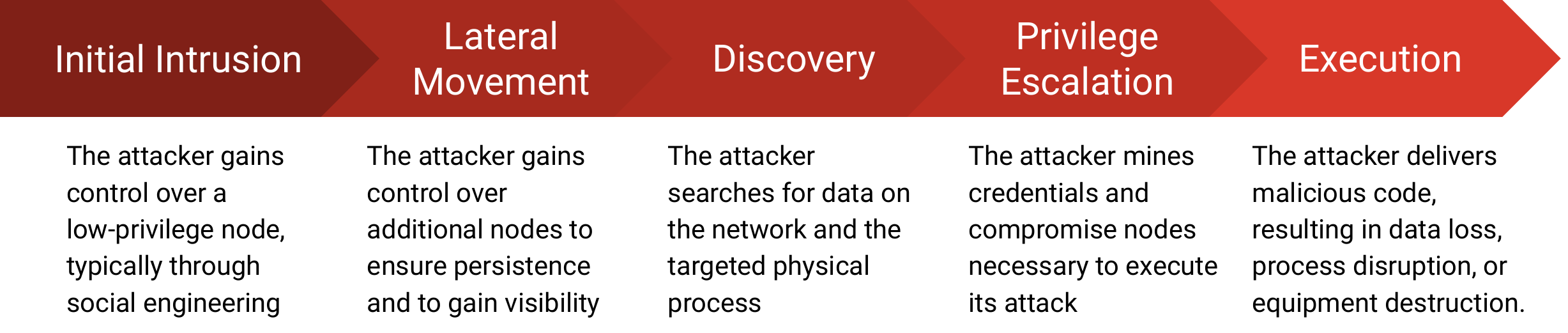}
    \caption{APT Attack Progression. This figure shows the typical progression of an APT attack at the level of tactics in the MITRE ATT\&CK framework. The process starts on the left with initial intrusion, proceeding over several months to eventual execution.}
    \label{fig:kill_chain}
\end{figure*}

Cyber attacks have been increasingly targeting computer networks that are integrated with industrial control systems (ICS)~\cite{das2020}.
Attack techniques focusing on stealth and redundant access make them difficult to detect and stop~\cite{alladi2020}. 
Intrusion detection systems monitor networks and alert security orchestrators to potential intrusions, though  high false-alarm rates cause many alerts to receive no security response. 
Advanced persistent threat (APT) attackers take advantage of this by spreading across target networks undetected over long periods to prepare an eventual attack~\cite{li2016}.
In addition to the theft of sensitive data common in APT attacks on standard networks, attacks on industrial control systems can additionally result in disruption of the controlled system, physical destruction of equipment, and even loss of life~\cite{langner2011, liang2016, di2018}.

It is not feasible for human security teams to respond to every potential intrusion alert and to reliably mitigate all threats.
This work seeks to demonstrate the feasibility of developing an automated cyber security orchestrator (ACSO) to assist human analysts by automatically investigating and mitigating potential attacks on ICS networks. 
We first present a model of this task as a discrete-time sequential decision making problem. 
The proposed model captures many of the challenges of the underlying decision problem while remaining agnostic to specific network configuration and communication dynamics.
To support this, we implemented a cyber attack simulator that can efficiently generate large quantities of trial data.

Many sequential decision making methods cannot solve this problem without explicit models of APT attack dynamics and alert behaviors that are not generally known.
Reinforcement learning (RL) can solve complex tasks without explicit models~\cite{vinyals2019, hoel2019}. 
There are several aspects of the computer network defense problem that make learning with existing reinforcement learning approaches difficult. 
The number of potential observations that an agent may make at a given time-step grows with the number of nodes on a network. 
Similarly, the number of mitigation actions the agent can take scales directly with the number of nodes. 
Deep reinforcement learning is known to struggle to learn on problems with large observation spaces~\cite{nguyen2020} and action spaces~\cite{dulac2015}.

APT attacks are designed to be difficult to detect, causing limited observability of the true compromise state of nodes on the network. 
Solving partially observable problems requires learning over sequences of observations to infer the hidden states~\cite{hausknecht2015}, making them significantly more difficult to solve than fully observable problems.
Effective defense against APTs requires timely response to compromises, though intrusion campaigns can last several months.
Modeling the decision process requires fine resolution time-steps, leading to very long time horizons.
Long time-horizons and sparse rewards can drastically increase the sample complexity of learning, both through exploration difficulty and temporally delayed credit assignment~\cite{andrychowicz2017, arjona-medina2019}.

This paper proposes a neural network architecture and training algorithm that scales to the large problem space.
We present an attention-based architecture that effectively learns over many cyber network nodes without growth in the number of required parameters. 
To overcome exploration difficulty, we introduce a potential-function shaping reward to aid training without biasing the converged policy behavior.
To simulate the alert generation with an intrusion detection system (IDS), we develop a dynamic Bayes network filter~\cite{normand1992} that aggregates network observations in an approximately optimal way. 

We tested the proposed solution methods against several baseline automation policies using the same simulation parameters that were used for training. 
To test the robustness of each approach to changes in attacker behavior, we evaluated performance with perturbations to APT attack policies and action effects.
aqThe results show that the reinforcement learning agent can successfully defend against attack with fewer disruptive mitigation actions than the baseline methods. 
Results also indicate that the agent is more robust to changes in attack trajectory than baselines.

\section{Background and Related Work} 


\begin{figure*}[bt]
    \centering
    \includegraphics[width=0.8\textwidth]{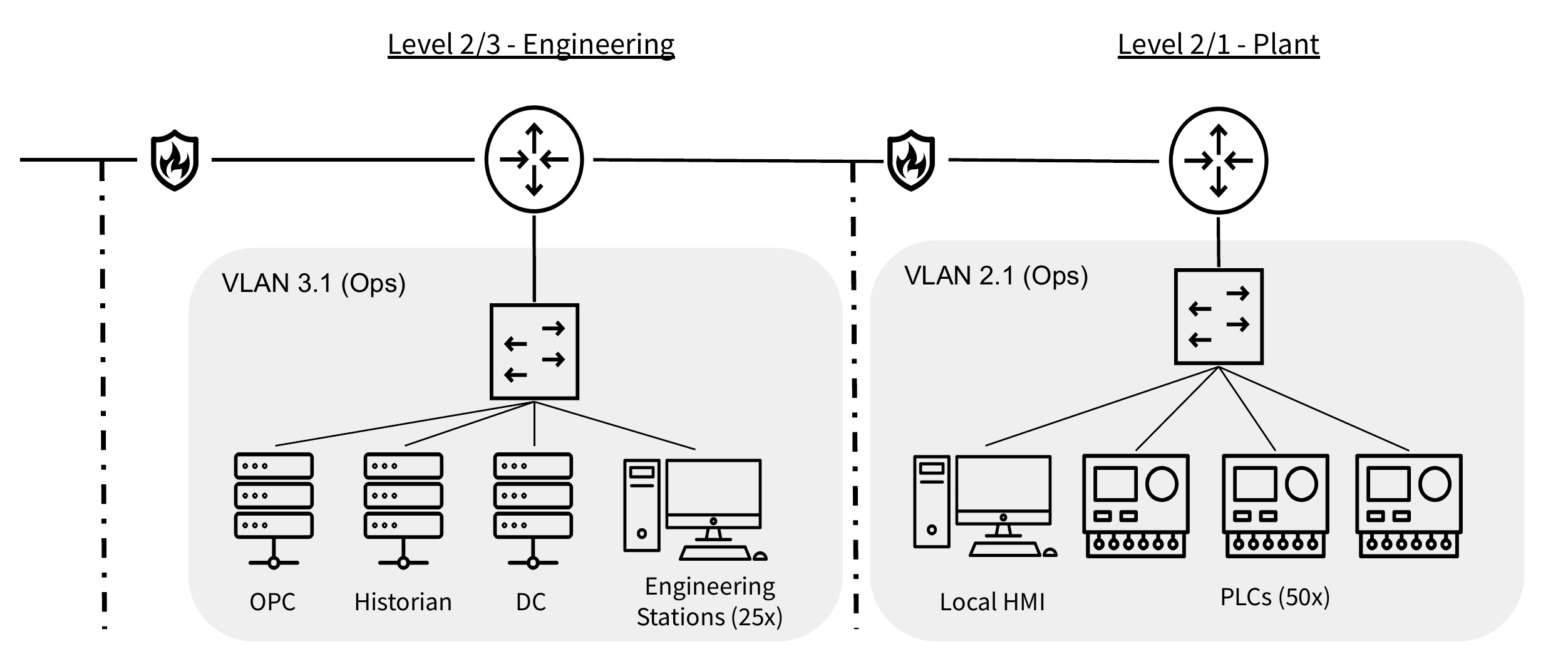}
    \caption{Simulated Network Architecture. The problem simulates level 2 and level 1 of the PERA model. Each level contains an operations VLAN and a nominally empty quarantine VLAN. Level 2 contains twenty-five workstation nodes and three servers. Level 1 has five local human-machine interface nodes and fifty networked PLCs. All network message traffic is simulated through virtual switches, routers, and firewalls. The computing node for the ACSO is not explicitly modeled.}
    \label{fig:network}
\end{figure*}

As their name implies, APTs are getting more sophisticated, and artificial intelligence (AI) offers growing potential for automating and providing intelligence to malware~\cite{kaloudi2020AI}.
APTs are increasingly targeting critical infrastructure systems.
These developments necessitate rapid advancements in cyber defense capabilities. 
In particular, active cyber defense (ACD) capabilities~\cite{herring2014active} need to mature beyond their reliance on domain expert developed courses of action (COA) in order to rapidly, and dynamically adapt to changing adversary tactics, techniques, and procedures (TTPs)~\cite{burke2020robust}. 
For improved cyber defense, AI has been proposed in a number of areas~\cite{apruzzese2018}:
\begin{itemize}
    \item Penetration Testing: Proactively identify vulnerabilities and patch them
    \item Detection and Identification: Use anomaly detection and pattern recognition to detect and classify malicious activity within a network
    \item Recovery and Response: Mitigate adversary actions and return the system to an operational state autonomously or in collaboration with a human operator
\end{itemize}

Every computer system has inherent vulnerabilities that an adversary can exploit. 
In order to identify and eliminate these weaknesses, system owners utilize teams of penetration testers to conduct exercises where they attempt to infiltrate the system, move laterally within the network, search for exploits and execute malicious payloads, as shown in the MITRE ATT\&CK framework in ~\cref{fig:kill_chain}. 
These exercises result in reports describing identified vulnerabilities which the system owner can then attempt to patch. 
In order to automate and speed up the process of identifying and patching vulnerabilities, DARPA conducted a Cyber Grand Challenge where teams developed AI-based systems to both probe for weaknesses in each other networks and patch vulnerabilities in their own networks~\cite{song2015darpa}.
For cyber defense, teams could either patch binaries or update IDS rules.
Xandra, which finished in second place, used fuzzing and symbolic execution to generate offensive proofs of vulnerability and mainly focused on binary patching for defense with a single, fixed network filter rule set that it selectively deployed~\cite{nguyen2018xandra}. 
Mayhem, which won the competition did not use the IDS system and only relied on binary patching~\cite{avgerinos2018mayhem}. 
These approaches did not extensively use machine learning and relied instead on pre-defined rules and procedures from software vulnerability analysis. 
Zennaro et al. propose an approach to us model-free reinforcement learning to solve capture the flag challenges, thus demonstrating an approach for automated penetration testing using machine learning~\cite{zennaro2020modeling}. 
While automated pen testing and patching capabilities are necessary in order to improve system robustness and resilience, their contributions to cyber defense are complementary to the automated attack mitigation described in this work.

The NIST cybersecurity framework defines five functions necessary for critical infrastructure system cybersecurity: Identify, Protect, Detect, Respond, and Recover~\cite{nist_2018}. 
A number of intrusion detection solutions, including Bro~\cite{paxson1999bro}, Snort~\cite{roesch1999snort}, and Security Onion~\cite{burks2012security} provide functionality such as identifying and detecting malicious activity. 
Companies such as Claroty~\cite{claroty} and Tenable~\cite{tenable} even produce IDSs focused on industrial control systems. 
Many of these solutions rely on pre-defined rules against which machine activity or network traffic is compared in order to flag anomalies.
Some recent solutions incorporate machine learning capabilities to develop models of normal network traffic which can then be used to identify anomalous behaviour. 
ACSO relies on inputs from IDSs such as these as well as other sensors deployed within computer networks in order to identify ongoing attacks. 
Due to the existing solutions and extensive research being conducted to address the first three phases of the NIST framework, ACSO is focused on enhancing decision making and responding to attacks in order to enable a compromised system to recover faster than would be possible with existing recovery and response capabilities.

Existing approaches to recovery and response rely on manned Security Operations Centers (SOCs) where human operators take actions in response to alerts generated by Security Information and Event Management Systems (SIEMs). 
Security Orchestration, Automation, and Response (SOAR) systems such as XSOAR and Splunk SOAR assist the cyber defenders by walking them through pre-defined COAs in response to alerts and indicators from network sensors. 
As Dhir et al. state, AI controlled attacks could simply overwhelm this current generation of cyber defense by adapting faster than pre-defined playbooks can respond~\cite{dhir2021prospective}. 
ACSO attempts to enhance the automation enabled by the current generation of SOAR technologies with reinforcement learning approaches.

As adversaries get more sophisticated and machine learning capabilities become more widely accessible, its likely that malware will be learning enabled. 
Such an advancement would result in an asymmetric environment where existing defensive approaches are reacting to malicious autonomous agents at human speed. 
In order to get ahead of this potentially catastrophic situation, it is necessary to develop capabilities such as the Autonomous Cyber Security Orchestrator (ACSO).

\section{Advanced Persistent Threat Scenario} 

\begin{figure*}[tb]
    \centering
    \includegraphics[width=0.75\textwidth]{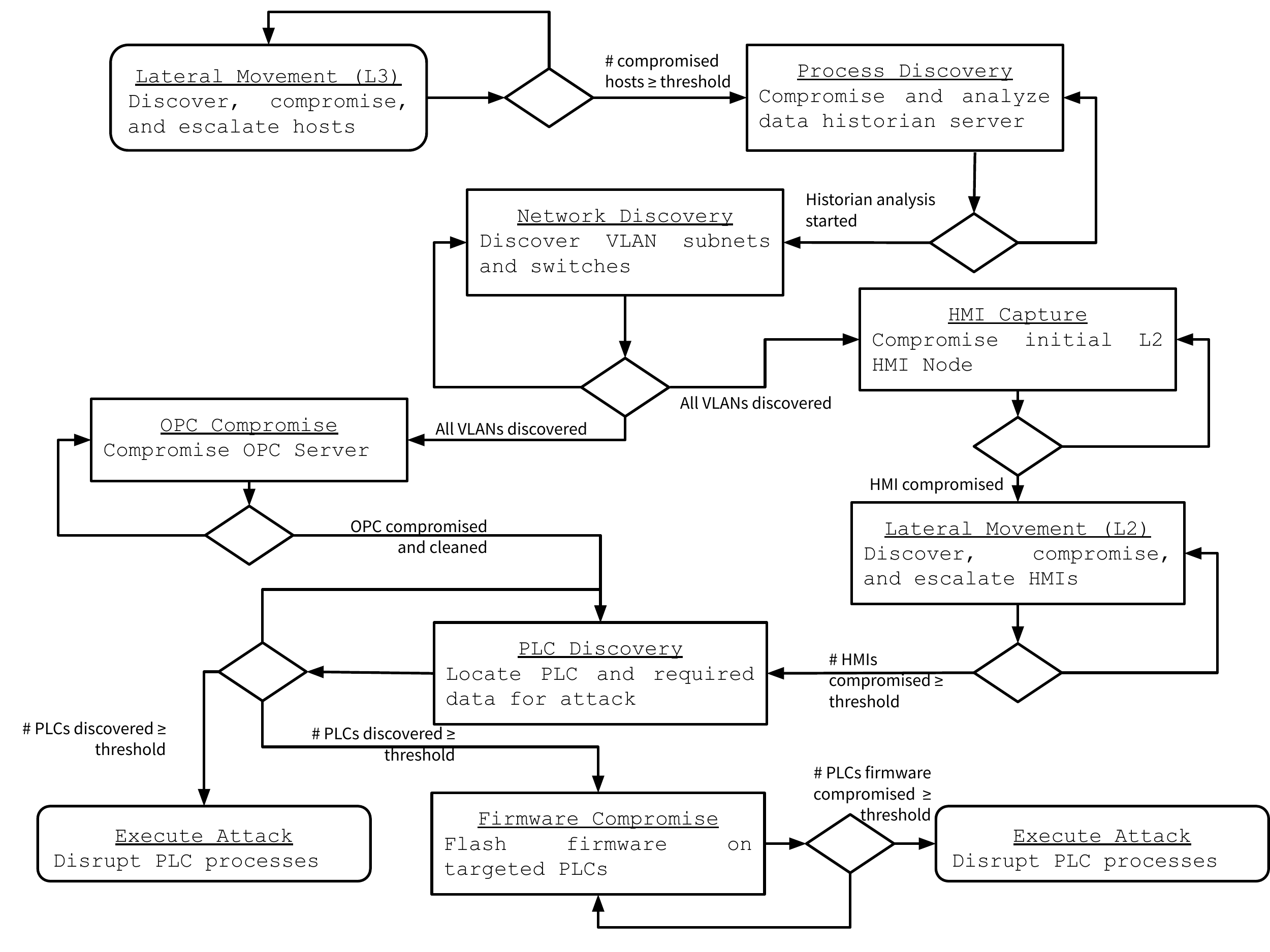}
    \caption{Attacker Trajectory. The simulated APT must meet various conditions to enact high-level tactics, moving towards the goal of disrupting or destroying PLCs. Rectangles indicate machine states that represent specific subroutines, diamonds represent exit criteria necessary for transitioning to the next state, and the rounded boxes indicate the start and end states.
    }
    \label{fig:attackgraph}
\end{figure*}

APTs tend to be well-funded and attack specific targets for significant periods of time~\cite{gosler2013}. 
APT attack goals vary from financial, to data theft, to physical infrastructure command and control. 
In this work, we focus on attacks aimed to disrupt or destroy industrial control systems.
APTs have been responsible for a number of recent, high-profile attacks that damaged or disabled infrastructure ICS. 
Attacks on the power grid in Ukraine and electrical utilities in the eastern United States have been linked to APT groups~\cite{lee2016analysis, eaton_volz_2021}. Additionally, a water treatment plant in Oldsmar, Florida was accessed through remote management software by a malicious actor attempting to alter chemical levels in the water ~\cite{peiser_2021}. 


ICS networks present unique defense difficulties.
The need to control equipment distributed across large geographical areas make internet-connected devices a mainstay of infrastructure systems.
To attempt to mitigate this vulnerability, many networks are organized into firewall-separated levels, with low-privilege nodes on less restricted and internet-accessible levels, and process-critical nodes on more isolated subnets~\cite{stouffer2011}.

APTs overcome the layered ICS network structure by first compromising a low-privilege node in the target network~\cite{meckl2017}.
Using this node, the APT will then conduct internal network reconnaissance and take control over more privileged nodes until it gains the authority needed to achieve its primary objective, for example to destroy equipment.
During this reconnaissance and lateral movement phase, intrusion detection systems tend to struggle to detect the attackers’ activity~\cite{li2016}.
This can extend over a period of months, while the attack itself is often executed in a matter of hours; thus, it is particularly important to disrupt the offensive during the staging phase.  

For this work the APT behavior is modeled at the level of tactics in the MITRE ATT\&CK framework~\cite{strom2016}.
Specific techniques, vulnerabilities, and exploits are not explicitly modeled.
The underlying assumption is that an APT will choose the right technique to achieve its tactical objective and the variance in time, odds of success, and alert probability can be represented by a stochastic distribution.
We model APT attacks that aim to capture programmable logic controllers (PLCs) at the plant level of the Purdue Enterprise Reference Architecture (PERA)~\cite{williams1993}. 

\Cref{fig:network} shows the network diagram used in this study, which is organized into levels 1 and 2 of the PERA architecture. 
Level 2 contains exclusively workstations (twenty-five) and servers (three), and level 1 contains five workstations and all fifty PLCs. 
We focus on attack scenarios that start with control of a beachhead node in level 2, from which the attacker’s goal is to disrupt or destroy the PLCs in level 1.
This scenario models the layered ICS network structure and APT attack approach described above, and thus represents a range of malicious activity that has impacted critical infrastructure, such as those on Ukraine's electrical grid.

\subsection{Simulation Environment} 
To develop and test our approach, we implemented an ICS network attack simulator (INASIM). 
INASIM runs high-level simulations of APT attacks on ICS networks in super-real time. 
Simulated networks are organized into levels as in the PERA architecture. 
The network is defined by three types of objects: nodes, networking devices, and PLCs. 
Nodes are computers that APTs may compromise to spread to other nodes and to launch attacks on PLCs.
Nodes may be workstation hosts or servers, and are connected to one-another through networking devices. 
The number of nodes, devices, PLCs, and the specific network connectivity, are all configurable in the simulator. 

The agent interacts with the environment through an API which defines the actions the agent can take and how these actions impact the state of the network. 
The API for this simulator is designed to provide a generic interface for interacting with networks of different sizes and connectivity.
Using the API, a defender agent may choose to take investigation or mitigation actions on any node at each decision step.
The result of an action taken on a node depends on that node's state.
Node states define how a node has been modified by the APT, either to enable greater APT permissions, to prevent detection, or to ensure persistence.
\Cref{tab: node states} lists the various compromise conditions that each node may experience. A node may have several compromise conditions at a given time. 
\begin{table*}[htbp]
    \centering
    \begin{tabular}{l l l}
         \toprule
         Condition & Effect & Required Condition  \\
         \midrule
         Scanned & Allows APT to gain command and control & None \\
         Initial Compromise & Allows APT to take actions on and from node & Scanned  \\
         Reboot Persistence & Prevents control loss from reboot & Initial compromise \\ 
         Admin Access & Enables additional actions & Initial compromise \\
         Credential Persistence & Prevents control loss on password change & Admin access \\
         Malware Cleaned & Reduces probability of alert generation & Admin access \\
         \bottomrule
    \end{tabular}
    \vspace{2mm}
    \caption{Node States. This table defines the possible compromise states a node may experience as a result of APT actions.}
    \label{tab: node states}
\end{table*}

Defender investigation actions do not change the states of nodes on the network, but instead stochastically provide alerts of the status of compromised nodes that are investigated. 
Mitigation actions change the states of nodes to impede the progress of APT attacks. 
Each action also has a cost defining relatively how much it disrupts nominal network operation. 
See the appendix for tables of available defender actions, along with each action's duration and cost, that were chosen to model the industrial control network security decision making problem. 

We model an Intrusion Detection System (IDS) that generates alerts based on activity in the network. 
Alerts provide the IP address of the node or networking device that generated the alert as well as a level of severity from 1 (lowest) to 3 (highest). 
Severity levels are based on the state of the node that generated the alert. 
Each action attempted by the APT may generate an alert from the IDS. 
Actions occurring only on a single node generate alerts with a probability given by the alert rate associated with an action. 
For actions that generate messages on the network, this rate is multiplied by the device factor for each networking device it passes through. 
The baseline IDS also passively generates alerts on compromised nodes with probability 0.1. 
Finally, the IDS generates false alerts at each level each hour of simulated time with $5*10^{-2}$ probability, $5*10^{-3}$ probability, and $2.5*10^{-3}$ probability for level 1, 2, and 3 severity respectively. 
See the appendix for full details on alert distributions.

\subsection{Attacker Model} 
Attacker behavior is generated from a stochastic finite state machine. 
Finite state machines are graphical models of intelligent behavior in which sub-routines are executed from a given machine state until certain exit conditions are met. 
The machine state then transitions to a new state and sub-routine until a final goal state is reached. 
The proposed finite state machine defines the high level actions available to the attacker based on the current state, similar to previous work modelling attack scenarios with graphs~\cite{panda2017near}.
The graph in Figure \ref{fig:attackgraph} shows the trajectories available to the APT at the tactic level. 
See the appendix for full details on actions available to the attacker.

The APT can either choose to access the PLCs by compromising the OPC server or the Level 1 HMI nodes. 
The APT can also choose whether to disrupt the ICS process or to destroy the plant equipment. 
Destroying equipment requires flashing PLC firmware, which is not required for PLC process disruption.
Because disruption does not require PLC firmware compromise, it is easier to achieve than PLC destruction, though the effects are less costly to the ICS than equipment destruction. 
Compromising the OPC server is easier as it only requires a single Level 2 node be compromised prior to attack. 
However, sending commands from Level 2 to Level 1 creates more alerts than sending commands from Level 1 HMIs. 
Each APT attack is additionally defined by several quantitative parameters.
The APT parameters are
\begin{itemize}
     \item Lateral Threshold: The number of Level 2 nodes to be compromised before escalating attack to the next phase. 
     \item PLC Threshold: The number of PLCs to discover before executing attack. 
     \item Labor budget: Each task takes some amount of time (in hours) and has an associated labor rate (in labor hours per clock hour). Each APT has a labor budget defining the maximum executable labor hours per wall clock hour.
\end{itemize}

The default APT policy has a lateral threshold of 3, and a PLC threshold of 15 for destroying attacks and 25 for disrupting attacks. 
The default policy assumes two full-time attackers at keyboard for a labor-rate of two. 

\section{Solution Method} 
In this work, we formulate network security as a sequential decision making problem. 
Sequential decision problems model the environment with states $s$ that evolve according to potentially stochastic dynamics. 
An agent takes actions $a$ that condition state transition distributions $T(s'\mid s, a)$ and generate rewards $r(s,a,s')$. 
In many problems, the state of the world is not known. 
In these partially observable domains, agents receive noisy observations according to $o \sim Z(o \mid s, a)$.

A sequential decision problem is solved by an action sequence that maximizes the expected value $ V(s) = \mathbb{E}\big[ \sum_t \gamma^t r(s_t, a_t)\big]$ for all states in the trajectory.  
Reinforcement learning methods learn a policy $\pi : o_{t-\tau:t} \mapsto a_t$ that maps a history of observations to actions through repeated trial-and-error with the environment. 
In each trial, actions are taken according to the current policy until a terminal condition is reached. 
At the end of the trial, the policy is updated to improve the expected performance.
Reinforcement learning methods that use neural networks to represent the learned policy are known as \emph{deep} reinforcement learning.

In this work, we solve a the simulated network security problem using deep reinforcement learning (DRL). 
We study reinforcement learning because it is a model-free method that does not require a formal, mathematical model of network dynamics to solve. 
Deep reinforcement learning with neural networks learns behavior that can generalize to situations not experienced in training. 
Reinforcement learning can train policies that are robust to mismatch between simulation and real-world dynamics and to adversarial approaches. 
Neural network policies can be continually updated from data observed during deployment. 
Unlike many formal game-theoretic methods, multi-object reinforcement learning can scale to very large networks~\cite{elderman2017adversarial}. 

We propose a learning method based on Deep Q-Network (DQN) learning. 
DQN is a popular method in which the learned neural network predicts the expected value $Q(o_{t-\tau:t}, a) = \mathbb{E}\big[r(s_t, a) + \gamma V(s_{t+1})\big]$ of taking each action in the action space for a given input history~\cite{mnih2013}. 
The Q-network is used as a policy by taking the action with the highest predicted value $a^* = \mathrm{argmax}_a Q(o_{t-\tau:t}, a)$.
The values of the network parameters are tuned with batches of experiences over episodes of the task. 

The network security problem presents many challenges to existing reinforcement learning methods, including high sample complexity, difficult exploration, and potentially adversarial dynamics. 
Deep RL training typically requires large amounts of data~\cite{wei2019}. 
The amount of trials required to solve a task tends to grow with the size of the input and output space of the problem. 
In tasks with very large input spaces, output spaces, or very long time horizons, the odds of finding a successful trajectory through random exploration are low~\cite{ecoffet2021}.
Adversarial dynamics emerge when an opposing agent in the environment observes the learned agent and adapts their own behavior to make the problem more difficult. 
In this work, we present methods to address the first two concerns and leave adversarial training to future work.  

The focus of this work is on making security decisions given some limited awareness of the network environment. 
This work does not propose methods to learn effective filtering of network observations, as this is a separate active area of research~\cite{apruzzese2018}. 
To avoid confounding the two problems, we implemented a dynamic Bayes network (DBN) filter as a surrogate for integrated network awareness systems.
The DBN learns approximate network dynamics and alert probabilities from data and provides a distribution over each node state at every time step.

\subsection{Problem Definition} 
The objective of the learning task is to prevent an attack with as  few disruptive actions as possible. 
To encode this objective into a suitable optimization target, we define a per-step reward function. 
The reward function $r(s, a, s')$ defines the "goodness" of taking action $a$ from state $s$ and then transitioning to state $s'$. 
The policy objective is to maximize the discounted sum of rewards over a time horizon. 

Reward each step is based on the fraction of PLCs that are operating nominally and the inconvenience caused by the ACSO actions taken. 
The reward function is defined as
\begin{align}~\label{eq: reward}
    r(s,a) &= r_\text{PLC}(s, a) + \lambda r_\text{IT}(s, a) + r_\text{term}(s,a) \\
    r_{PLC}(s,a) &= \Big(1 - 0.05 n_{disrupted} - 0.1 n_{destroyed} \Big) \\ 
    r_{IT}(s,a) &= \Big(1 - \sum_{a \in A_t}\mathrm{cost}(a)\Big)\\
    r_{term}(s,a) &= \frac{1}{1 - \gamma}1\{s_\text{time} \geq t_\text{max}\}
\end{align}
where $\lambda$ is a weighting parameter and $n_{disrupted}$ and $n_{destroyed}$ give the total number of disrupted and destroyed PLCs, respectively. 
$A_t$ is the set of all actions completing at time $t$.

The first term $r_{PLC}$ rewards the agent for preventing the PLCs from being compromised. 
The second term imposes a penalty for ACSO actions, where each action is assigned a cost based on its perceived burden to network operations.
The first and second terms encode opposing objectives. 
To prioritize PLC defense, the $\lambda$ weighting term can be set to less than 1. 
In this work, $\lambda = 0.1$ for all experiments. 
The final term rewards the agent for reaching the episode time limit $t_{\max}$.
The $1/(1 - \gamma)$ magnitude of the terminal reward ensures the optimal state value does not drift with episode time. 

Actions that are more effective at securing compromised nodes are given a higher cost. 
For example the low-disruption action of rebooting a workstation has a cost of 0.01, while the more disruptive action of re-imaging a server has a cost of 0.05.
Action costs were set to provide a range of high, medium,  and low disruption actions. 

Each time step of the simulation corresponds to one hour of real-world time. 
We run each trial episode for a maximum of 5,000 time steps or approximately six months. 
This was chosen to ensure the attacker would have enough time to complete an intrusion campaign. 
The time discount factor was set to $\gamma = 0.9995$. 
Having a discount rate near 1 allows the agent to learn to account for action effects far into the future. 
Given this discount rate, the maximum discounted return for a given episode is $2200$, though achieving this would require defending the network without taking any actions. 

\subsection{Training Algorithm} 
We train our neural network to estimate the value of taking each available action $a$ given a history of previous observations on the network $h$.
The proposed solution uses an augmented DQN algorithm to learn a policy defined by our attention-based neural network.
We implemented several extensions to the baseline algorithm, based on studies of Rainbow DQN~\cite{hessel2018}.
The included extensions are double-DQN~\cite{hasselt2010}, prioritized experience replay~\cite{schaul2016}, and $n$-step TD loss~\cite{sutton1998}.

The policy is trained to minimize the error in its estimation, defined by a temporal difference (TD) loss. 
The TD loss uses a bootstrap estimation method to reduce training sample variance and a second copy of the policy network, called the target network, to reduce sample over-estimation.
The training loss for a given step is 
\begin{align}~\label{eq: td_loss}
    \Big\| \Big(\sum_{\tau=t+1}^{t+n} \gamma^{\tau - t}r_\tau + \gamma^nQ_\phi\big(h_{t+n}, a_{t+n}\big)\Big) - Q_\pi(h_t, a) \Big\|
\end{align}
where $h_t$ is the history of observations at time $t$, $Q_\pi$ is the action value estimate of the policy network and $Q_\phi$ is the action value estimate of the target network. 
The action $a_{t+n}$ is given as $\mathrm{argmax}_{a'} Q_\pi(h_{t+n}, a')$.
The $\| \cdot \|$ represents the Huber-loss norm. 
This loss is calculated over batches of importance-weighted samples from an experience replay buffer and used to estimate the gradient for each network update. 

In addition to the task reward presented in~\cref{eq: reward}, a shaping reward was defined based on the non-biased potential formulation~\cite{ng1999}.
This reward was designed to incentivize the agent to secure compromised nodes. 
The shaping function was defined as  
\begin{equation}~\label{eq: train_reward}
    r_{shape}(s,a,s') = \gamma(A\delta_{W} + B\delta_{S})
\end{equation}
where $\delta_{W}$ and $\delta_{S}$ are changes in the number of workstations and servers compromised by the APT from state $s$ to $s'$, respectively, and $A$ and $B$ are weight factors.
The weighted sum of~\cref{eq: reward} and~\cref{eq: train_reward} were used for training.
Only~\cref{eq: reward} was used for evaluation.
This additional shaping reward was critical to enable the agent to learn a meaningful policy, as the baseline reward function provided too sparse of a signal over the long episode lengths.

Training hyper-parameters were tuned by a grid search training on a smaller ICS network with ten level 2 workstation nodes, three level 1 workstations nodes, and thirty PLCs. 
We searched over the shaping reward weight, the observation-history interval, the target network update frequency, and the $\epsilon$-greedy exploration decay schedule. 
The parameter set leading to the highest average performing policy over several seeds after 500 episodes was selected. 
We used the PyTorch framework for neural network implementation and training~\cite{paszke2015}.
Numerical values for the training parameters as well as additional practical details on the training process can be found in the appendix.

\subsection{Dynamic Bayes Network}
\begin{figure}[htbp]
    \centering
    \includegraphics[width=0.7\columnwidth]{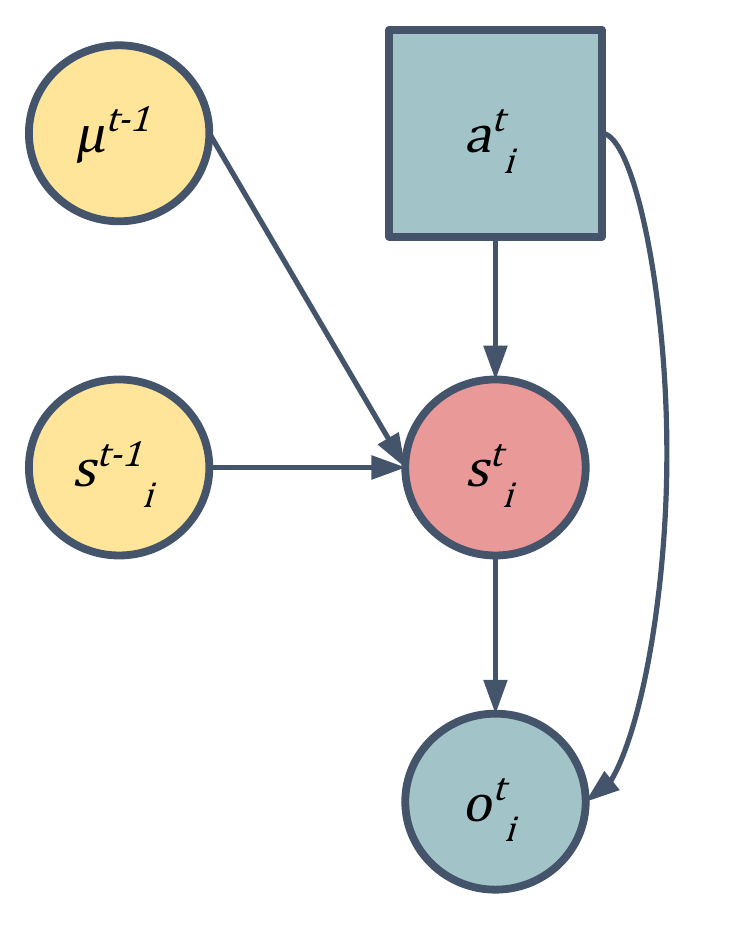}
    \caption{Dynamic Bayes Network. This figure shows the DBN filter for node $s_i$. Circular nodes denote random variables each with an associated conditional probability distribution, and square nodes represent actions. The probability distribution associated with a random variable is conditioned on the values of that variable's parent nodes, indicated by the arrow direction. Values for yellow nodes $\mu^{t-1}$ and $s^{t-1}$ are calculated from previous beliefs. The cyan nodes $a^t$ and $o^t$ represent values known during calculation. The probability distribution associated with the variable $s_t$ is the updated belief.}
    \label{fig:dbn}
\end{figure}

The network simulation generates alerts of potential compromise on nodes at each time step. 
The perception problem of inferring the compromise state from sequences of these observations is not the focus of this work.
To filter observations we implemented a dynamic Bayes network (DBN). 
The DBN takes observations $o_t$ from the network and produces a distribution over the compromise state of each node.
We refer to the distribution over possible states as a belief.
A graphical depiction of the DBN is shown in~\cref{fig:dbn}. 

\begin{figure*}[ht]
    \centering
    \includegraphics[width=0.75\textwidth]{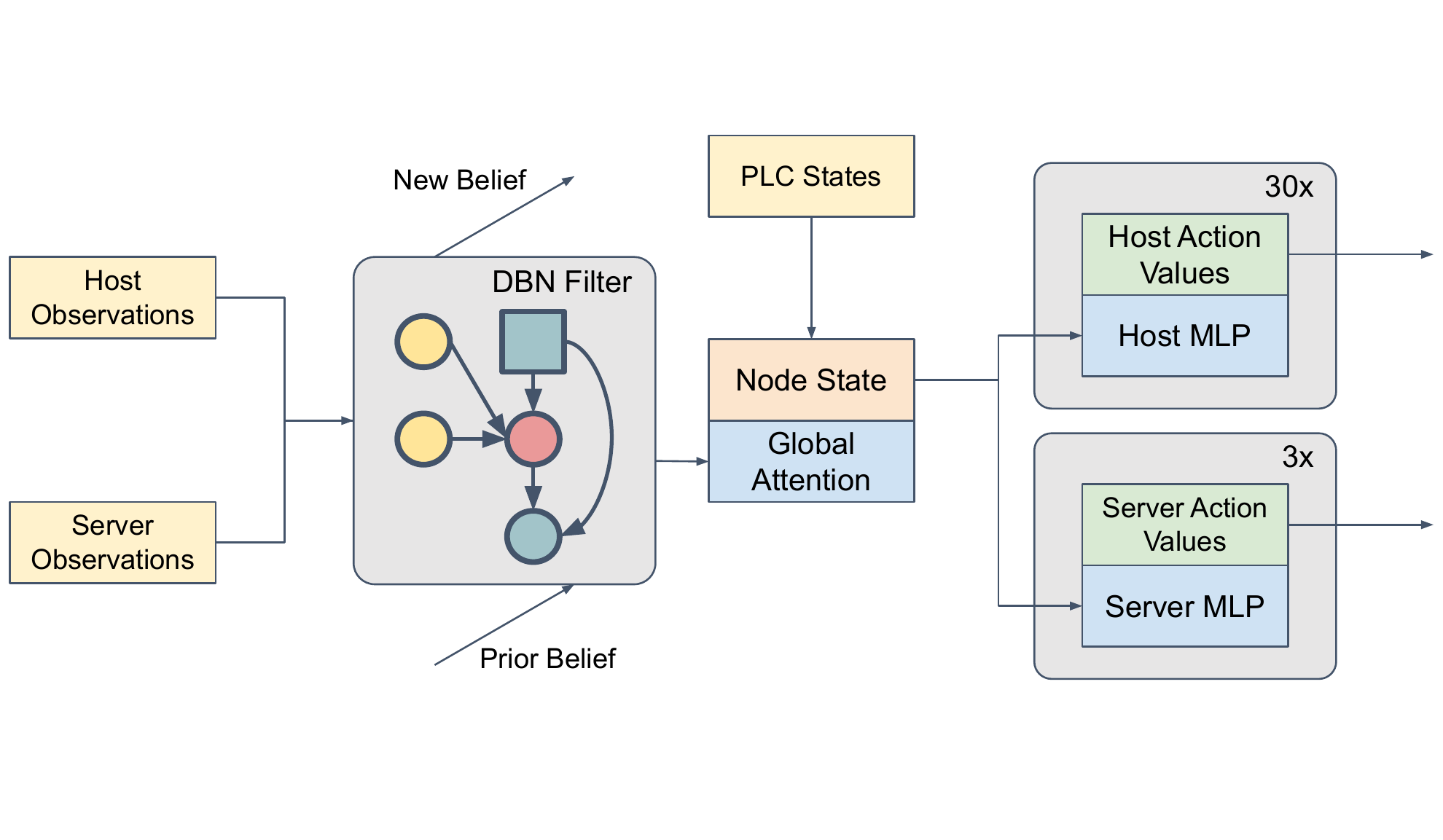}
    \caption{Neural Network. Inputs for each network node are passed into the dynamic Bayes network filter. The DBN updates the prior belief of each node's compromise with the new action and observation. These vectors are stacked and passed through a self-attention sub-graph, to provide global context to each latent vector. These vectors are then passed through fully connected sub-graphs to output action value estimates for each node.}
    \label{fig:neural_network}
\end{figure*}

The belief of a node $i$ is represented as a vector $b_i$, where each element gives the probability of that node having been compromised in a particular way.  
The DBN calculates beliefs by recursively applying Bayes rule. 
Given a belief on a node state $s_i^{t-1}$, the new belief after having taken action $a_i^t$ and observing $o_i^t$ is calculated as
\begin{equation}~\label{eq: dbn update}
    \eta P(o_i^t \mid s_i^t, a_i^t) \sum_{s_i^{t-1}\in \mathcal{S}} P(s_i^{t} \mid s_i^{t-1}, \mu_{t-1}, a_i^t)b_i^{t-1}(s_i^{t-1})
\end{equation}
where $\eta$ is a normalizing constant and $\mu_{t-1}$ is a summary statistic on the complete network state. 

The update in~\cref{eq: dbn update} approximates the optimal Bayesian update. 
The true optimal update conditions the transition probability of a node on the state of every other node in the network. 
Calculating this would require enumerating over every possible combination of node states and is intractable for networks with more than a small number of nodes. 
Instead, we condition on summary values, such as the total number of compromised nodes. 
These values are represented by the vector $\mu$.

The conditional probability distributions are not generally known. 
We learned these distributions from data. 
We ran 1,000 episodes with the APT and a defender policy that takes random actions each step.
For each episode we recorded the states, actions, and observations at each step. 
Using these values, we calculated probability tables for each distribution.
We validated the DBN performance by measuring the maximum Kullback-Leibler (KL) divergence~\cite{kullback1951} of the DBN belief and the true state over many episodes.

\subsection{Neural Network Policy} 

The ACSO problem has input and output spaces with dimensions that scale with the number of nodes in the ICS network.
Specialized neural architectures can take advantage of structure inherent in input data to improve training efficiency.
For example, convolutional nets leverage spatial invariance in images and recurrent networks encode correlation in sequences~\cite{krizhevsky2012, cho2014}.
We designed the neural network shown in~\cref{fig:neural_network} using attention mechanisms~\cite{vaswani2017} to accommodate the large input space without significant degradation of explanatory capacity or intractable growth in size. 
Attention mechanisms have been shown to improve learning efficiency on tasks with exchangeable inputs~\cite{mern2019, mern2020}.

%
Each node observation is passed through the DBN and a belief vector is returned. 
The belief vectors are then stacked and input to a global attention sub-graph. 
This sub-graph allows the network to learn which features of neighboring nodes are relevant to the value function of actions on a given node. 
In this problem, we assume that we can directly observe whether or not a PLC process has been successfully disrupted or equipment destroyed. 
A vector encoding the PLC states is concatenated with the node belief vectors. 
These contextualized node vectors are then passed to feed-forward output sub-graphs. 
All sub-graphs of a given node type share the same parameter set, so a growth in the number of nodes does not cause a growth in the total number of network parameters. 



\section{Experiments} 
\begin{table*}[htbp]
    \centering
    \begin{tabular}{l c c c c}
         \toprule
         Policy & Discounted Return & Final PLCs Offline & Average IT Cost & Average Nodes Compromised\\
         \midrule
         ACSO & $\mathbf{2149.9 \pm 0.2}$ & $\mathbf{0.0 \pm 0.0}$ & $\mathbf{0.15 \pm 0.0}$ & $\mathbf{0.56 \pm 0.0}$ \\
         DBN Expert & $1970.5 \pm 26.6$ & $5.6 \pm 0.9$ & $0.40 \pm 0.0$ & $0.62 \pm 0.0$ \\
         Playbook & $2142.6 \pm 0.1$ & $\mathbf{0.0 \pm 0.0}$ & $0.21 \pm 0.0$ & $0.63 \pm 0.1$ \\
         Semi Random & $2071.9 \pm 0.1$ & $\mathbf{0.0 \pm 0.0}$ & $0.60 \pm 0.0$ & $0.88 \pm 0.6$ \\
         \bottomrule
    \end{tabular}
    \vspace{2mm}
    \caption{Nominal Evaluation Results. 
    This table presents the performance of the ACSO agent and the three baseline policies tested with nominal simulation parameters. 
    The discounted task return, total number of PLCs offline, average disruptive IT cost per step, and the average number of nodes compromised by the APT per hour are reported. 
    For each metric, the mean and one standard error bounds over 100 episodes are given.
    }
    \label{tab:performance}
\end{table*}

We tested the ability of the trained ACSO to defend against simulated attacks. 
We ran trials with the same simulation parameters used during policy training. 
To test policy robustness, we also tested performance under several deviations to APT behavior. 
We also implemented and evaluated several baseline policies for comparison.
For each test, we ran 100 attack scenarios and measured performance using three metrics: 
\begin{itemize}
\item PLCs offline: the total number of PLCs disrupted or destroyed during the attack.
Minimizing this is the main objective of a defender agent.

\item Average IT cost: the total disruption to nominal IT network behavior caused by the ACSO actions.

Policies that take actions more aggressively will incur a higher average action cost.

\item Average nodes compromised per hour: how many total nodes in levels 1 and 2 that the APT has command and control on in the network.
This metric gives a holistic measure of the total network compromise during the episode. 
\end{itemize}

We perturbed the APT policy in two ways. 
First, we changed how effective the APT was at evading detection. 
To do this, we modified the APT cleanup effectiveness parameter that made its node cleanup actions more effective at preventing alerts.
Higher cleanup effectiveness denotes a lower probability of detection.
We tested policy performance with cleanup effectiveness rates both higher and lower than the nominal rate used in training. 

Our second APT perturbation modified the qualitative parameters to create a more aggressive attacker. 
The number of lateral 3 hosts controlled before beginning process discovery was changed from three to one.
The number of PLCs compromised required to execute the attack was changed from fifteen to five when destroying equipment and from twenty-five to ten when disrupting. 
The result of these changes is an attacker model, referred to as APT2, that moves more quickly through the tactics graph shown in \cref{fig:attackgraph}, while being less resilient to setbacks, as the APT has less redundant access.

\subsection{Baseline Policies} 

We implemented three different baseline policies to compare to the ACSO. 
A random policy simulates independent security analysts and users taking actions on the network. 
The random policy takes actions by sampling action type from a static categorical distribution and a node uniformly from the nodes of the appropriate type in the network.
A playbook policy executes pre-defined COAs when specific alert conditions are met. 
Finally, an expert policy takes actions stochastically conditioned on the compromise predictions of the dynamic Bayes net.

The security automation playbook baseline defines fixed responses to alerts. 
For the scope of defender actions and IDS alerts that we consider, it is infeasible to create playbooks that address the entire decision making space. 
We only design playbooks that are triggered by a single alert, of which there are 3 types: An alert triggered by an APT action, a passive alert generated on a host node, and a passive alert generated on a server node. 
See the appendix for figures of the playbook courses of action triggered by a passive host alert. 
This playbook baseline is more automated and therefore provides faster response than most playbooks used in practice, as most security playbooks today defer to human analysts at crucial decision points. 

The expert policy samples actions from a distribution conditioned on the output of the DBN filter. 
The DBN estimates the compromise state of each node in the network from the observed alerts, and the most appropriate action is chosen based on the believed state of a node. 
For example, if a node is believed to be compromised, with no reboot persistence, then a reboot action will be taken, and if a node is believed to be compromised with credential persistence, a re-image action will be taken.

\section{Results}~\label{sec: results} 

\begin{figure*}[htbp]
     \centering
     \begin{subfigure}[b]{0.45\textwidth}
         \centering
         \includegraphics[width=\textwidth]{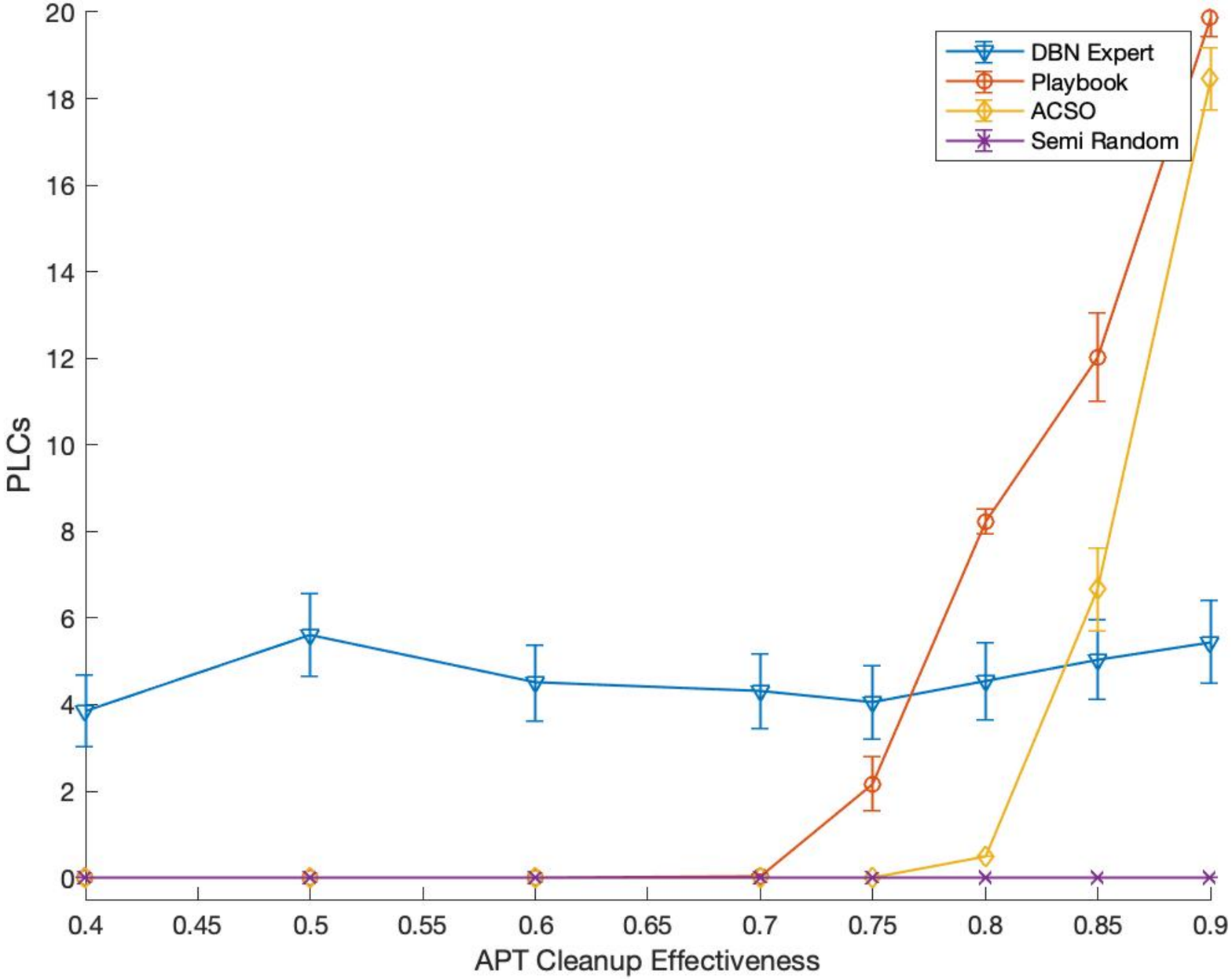}
         \caption{Final PLCs Offline}
         \label{fig:plcs}
     \end{subfigure}
     \begin{subfigure}[b]{0.45\textwidth}
         \centering
         \includegraphics[width=\textwidth]{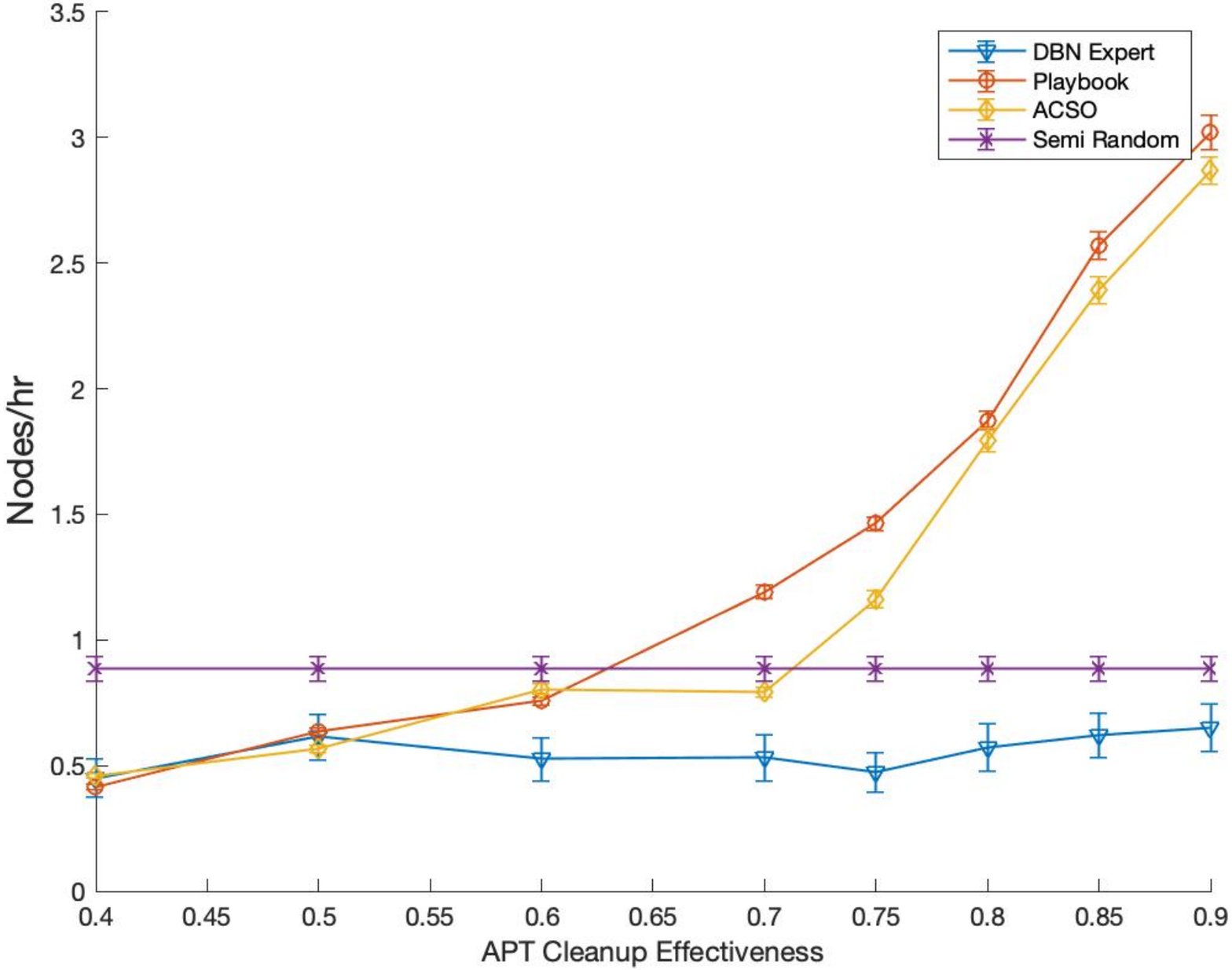}
         \caption{Average Layer 2/1 Nodes Compromised}
         \label{fig:nodes}
     \end{subfigure}
        \caption{APT Cleanup Effectiveness Experiments. These figures show the performance of the defender policies under perturbations to the APT cleanup effectiveness. APT cleanup effectiveness modifies the probability of a scan detecting the APT after the APT removed malware from the node. An effectiveness of 0.5 denotes that detections are 0.5 as likely after cleanup. The ACSO was trained with an APT cleanup effectiveness of 0.5.}
        \label{fig:cleanup rate}
\end{figure*}

The results from the experiments with the baseline environment are shown in~\cref{tab:performance}.
The learned ACSO policy performance meets or exceeds the performance of every baseline policy in all four metrics. 
Both the playbook and the ACSO policies greatly outperformed the random baseline, suggesting that coordinated, automated response significantly improves network security over random, decentralized behavior.
All policies, except for the DBN expert were able to successfully prevent any PLCs from being attacked, however, only the ACSO and playbook did so with average network disruption under 0.25. 
The ACSO was the most efficient defender, with an average IT disruption rate nearly $30\%$ lower than the disruption rate of the playbook. 
The ACSO agent was also the most effective at securing compromised network nodes, with just $0.56$ nodes compromised per-hour, which is $12\%$ lower than the compromise rate of the playbook and $37\%$ lower than the random rate. 

The performance of the policies under perturbations of the APT evasion effectiveness are shown in~\cref{fig:cleanup rate}.
As can be seen, as the cleanup effectiveness increases from the nominal $0.5$ level, both the ACSO and playbook eventually begin to fail. 
The playbook fails sooner and more sharply than the ACSO. 
Additionally, the ACSO policy maintains a lower action cost and lower average nodes compromised for most cleanup rates. 
The DBN policy is not sensitive to the change in perception characteristics in the environment, and uses a more aggressive approach, leading to a significantly higher action cost than the other policies. 
These results show that while ACSO was not trained in an environment with different detection probabilities after APT cleanup, it is robust to moderate changes.

The results from experiments done with a less cautious attack policy demonstrate that ACSO is more robust to changes in APT behavior than other policies we evaluated. 
See~\cref{fig:apt2} in the appendix for full results.
The ACSO performance in this setting is similar to its performance against the nominal attacker policy.
The ACSO is able to fully prevent PLC compromise while using the least disruptive actions of any of the tested policies with an average IT cost of 0.149, a 31\% reduction when compared to the playbook policy, which had the next best average IT cost of 0.216. 
The more aggressive attacker was able to compromise PLCs in a number of episodes when tested against the playbook policy, causing an average of 0.45 PLCs to be offline at the end of an episode. 
Both the DBN-based policy and random policy are slightly more effective against APT2 compared to the APT1, and the policies' average action costs suggest that they maintain a fairly aggressive stance towards defending the network. 

These results showing greater robustness of the ACSO are partially due to ability of the neural network abstractions to generalize learned behavior to new experiences. 


\section{Conclusions} 

In this work, we demonstrate the feasibility of using deep reinforcement learning to develop agents to autonomously secure a computer network.
We presented a high-level simulation environment able to efficiently generate the large amounts of trial data needed for RL training. 
We proposed a solution method to overcome several of the challenges that the ACSO learning problem poses to conventional deep RL. 
The dynamic Bayes network filter approximated an optimal observation filter, allowing this study to focus on the decision making problem without simultaneously learning a perception system.
The attention-based neural network allowed the learned agent to scale to large, varying-sized computer networks without an intractable growth in parameters. 
We proposed a shaping reward which allowed the agent to learn over the long episode lengths without biasing the final solutions. 

We tested the ACSO performance against baseline policies in several experiments.
Results show that the ACSO is able to outperform baseline policies when tested in the same environment in which it was trained. 
The results also show that the ACSO is more robust to changes in the APT behavior than a comparable rules-based playbook policy, despite not having been trained with examples of that behavior.

This work is an initial study of deep reinforcement learning in holistic computer network defense. 
Future work should prioritize improving the testing and training infrastructure, for example by improving the fidelity and flexibility of simulations. 
Data-efficient methods to validate learned policies performance in high-fidelity emulations should be developed.
Methods for pre-training models using simulations, and fine-tuning for deployment to specific ICS networks should be explored.
This work only considered a narrow set of attacker behaviors modeling current, known attack trajectories. 
As attackers continue to evolve and incorporate learning and planning in their attacks, focus should be placed on adversarial learning methods that can discover and obviate new attacks before they are observed in the real-world.

\section*{Acknowledgment}
This material is based upon work supported by the Johns Hopkins University Applied Physics Laboratory, as well as the Institute for Assured Autonomy.
The work was also supported by the Stanford University Institute for Human-Centered AI (HAI) and Google Cloud. 

\section*{Availability}
The code for the network simulation environment and APT policies are available as a Julia repository. 
The code required to train and test the neural network agent are also available. 
Scripts to run all experiments presented are provided. 
Links to code will be made public after anonymous review. 
\bibliographystyle{./bibliography/IEEEtran}
\bibliography{./bibliography/bib}

\begin{thebibliography}{10}
\providecommand{\url}[1]{#1}
\csname url@samestyle\endcsname
\providecommand{\newblock}{\relax}
\providecommand{\bibinfo}[2]{#2}
\providecommand{\BIBentrySTDinterwordspacing}{\spaceskip=0pt\relax}
\providecommand{\BIBentryALTinterwordstretchfactor}{4}
\providecommand{\BIBentryALTinterwordspacing}{\spaceskip=\fontdimen2\font plus
\BIBentryALTinterwordstretchfactor\fontdimen3\font minus
  \fontdimen4\font\relax}
\providecommand{\BIBforeignlanguage}[2]{{%
\expandafter\ifx\csname l@#1\endcsname\relax
\typeout{** WARNING: IEEEtran.bst: No hyphenation pattern has been}%
\typeout{** loaded for the language `#1'. Using the pattern for}%
\typeout{** the default language instead.}%
\else
\language=\csname l@#1\endcsname
\fi
#2}}
\providecommand{\BIBdecl}{\relax}
\BIBdecl

\bibitem{das2020}
R.~Das and M.~Z. G{\"u}nd{\"u}z, ``Analysis of cyber-attacks in {I}o{T}-based
  critical infrastructures,'' \emph{International Journal of Information
  Security Science}, vol.~8, no.~4, pp. 122--133, 2020.

\bibitem{alladi2020}
T.~Alladi, V.~Chamola, and S.~Zeadally, ``Industrial control systems:
  Cyberattack trends and countermeasures,'' \emph{Computer Communications},
  vol. 155, pp. 1--8, 2020.

\bibitem{li2016}
M.~Li, W.~Huang, Y.~Wang, W.~Fan, and J.~Li, ``The study of {APT} attack stage
  model,'' in \emph{IEEE/ACIS International Conference on Computer and
  Information Science (ICIS)}, vol.~15, 2016, pp. 1--5.

\bibitem{langner2011}
R.~Langner, ``Stuxnet: Dissecting a cyberwarfare weapon,'' \emph{{IEEE}
  Security and Privacy}, vol.~9, no.~3, pp. 49--51, 2011.

\bibitem{liang2016}
G.~Liang, S.~R. Weller, J.~Zhao, F.~Luo, and Z.~Y. Dong, ``The 2015 {U}kraine
  blackout: Implications for false data injection attacks,'' \emph{IEEE
  Transactions on Power Systems}, vol.~32, no.~4, pp. 3317--3318, 2016.

\bibitem{di2018}
A.~Di~Pinto, Y.~Dragoni, and A.~Carcano, ``{TRITON}: The first {ICS} cyber
  attack on safety instrument systems,'' in \emph{Black Hat USA}, 2018, pp.
  1--26.

\bibitem{vinyals2019}
O.~Vinyals, I.~Babuschkin, W.~M. Czarnecki, M.~Mathieu, A.~Dudzik, J.~Chung,
  D.~H. Choi, R.~Powell, T.~Ewalds, P.~Georgiev, J.~Oh, D.~Horgan, M.~Kroiss,
  I.~Danihelka, A.~Huang, L.~Sifre, T.~Cai, J.~P. Agapiou, M.~Jaderberg, A.~S.
  Vezhnevets, R.~Leblond, T.~Pohlen, V.~Dalibard, D.~Budden, Y.~Sulsky,
  J.~Molloy, T.~L. Paine, {\c{C}}.~G{\"{u}}l{\c{c}}ehre, Z.~Wang, T.~Pfaff,
  Y.~Wu, R.~Ring, D.~Yogatama, D.~W{\"{u}}nsch, K.~McKinney, O.~Smith,
  T.~Schaul, T.~P. Lillicrap, K.~Kavukcuoglu, D.~Hassabis, C.~Apps, and
  D.~Silver, ``Grandmaster level in {S}tar{C}raft {II} using multi-agent
  reinforcement learning,'' \emph{Nature}, vol. 575, no. 7782, pp. 350--354,
  2019.

\bibitem{hoel2019}
C.-J. Hoel, K.~Driggs-Campbell, K.~Wolff, L.~Laine, and M.~J. Kochenderfer,
  ``Combining planning and deep reinforcement learning in tactical decision
  making for autonomous driving,'' \emph{IEEE Transactions on Intelligent
  Vehicles}, vol.~5, no.~2, pp. 294--305, 2019.

\bibitem{nguyen2020}
T.~T. Nguyen, N.~D. Nguyen, and S.~Nahavandi, ``Deep reinforcement learning for
  multiagent systems: {A} review of challenges, solutions, and applications,''
  \emph{{IEEE} Transactions on Cybernetics}, vol.~50, no.~9, pp. 3826--3839,
  2020.

\bibitem{dulac2015}
G.~Dulac{-}Arnold, R.~Evans, P.~Sunehag, and B.~Coppin, ``Reinforcement
  learning in large discrete action spaces,'' \emph{Computing Research
  Repository}, 2015.

\bibitem{hausknecht2015}
M.~J. Hausknecht and P.~Stone, ``Deep recurrent {Q}-learning for partially
  observable {MDP}s,'' in \emph{AAAI Conference on Artificial Intelligence
  (AAAI)}, 2015, pp. 29--37.

\bibitem{andrychowicz2017}
M.~Andrychowicz, D.~Crow, A.~Ray, J.~Schneider, R.~Fong, P.~Welinder,
  B.~McGrew, J.~Tobin, P.~Abbeel, and W.~Zaremba, ``Hindsight experience
  replay,'' in \emph{Advances in Neural Information Processing Systems
  (NeurIPS)}, 2017, pp. 5048--5058.

\bibitem{arjona-medina2019}
J.~A. Arjona{-}Medina, M.~Gillhofer, M.~Widrich, T.~Unterthiner,
  J.~Brandstetter, and S.~Hochreiter, ``{RUDDER:} return decomposition for
  delayed rewards,'' in \emph{Advances in Neural Information Processing Systems
  (NeurIPS)}, 2019, pp. 13\,544--13\,555.

\bibitem{normand1992}
S.-U. Normand and D.~Tritchler, ``Parameter updating in a {B}ayes network,''
  \emph{Journal of the American Statistical Association}, 1992.

\bibitem{kaloudi2020AI}
N.~Kaloudi and J.~Li, ``The ai-based cyber threat landscape: A survey,''
  \emph{ACM Computing Surveys (CSUR)}, vol.~53, no.~1, pp. 1--34, 2020.

\bibitem{herring2014active}
M.~J. Herring and K.~D. Willett, ``Active cyber defense: a vision for real-time
  cyber defense,'' \emph{Journal of Information Warfare}, vol.~13, no.~2, pp.
  46--55, 2014.

\bibitem{burke2020robust}
A.~Burke, ``Robust artificial intelligence for active cyber defence,''
  \emph{Alan Turing Institute, Tech. Rep}, 2020.

\bibitem{apruzzese2018}
G.~Apruzzese, M.~Colajanni, L.~Ferretti, A.~Guido, and M.~Marchetti, ``On the
  effectiveness of machine and deep learning for cyber security,'' in
  \emph{IEEE Conference on Cyber Conflict (CyCon)}, 2018, pp. 371--390.

\bibitem{song2015darpa}
J.~Song and J.~Alves-Foss, ``The {DARPA} cyber grand challenge: A competitor's
  perspective,'' \emph{IEEE Security \& Privacy}, vol.~13, no.~6, pp. 72--76,
  2015.

\bibitem{nguyen2018xandra}
A.~Nguyen-Tuong, D.~Melski, J.~W. Davidson, M.~Co, W.~Hawkins, J.~D. Hiser,
  D.~Morris, D.~Nguyen, and E.~Rizzi, ``Xandra: An autonomous cyber battle
  system for the cyber grand challenge,'' \emph{IEEE Security \& Privacy},
  vol.~16, no.~2, pp. 42--51, 2018.

\bibitem{avgerinos2018mayhem}
T.~Avgerinos, D.~Brumley, J.~Davis, R.~Goulden, T.~Nighswander, A.~Rebert, and
  N.~Williamson, ``The {M}ayhem cyber reasoning system,'' \emph{IEEE Security
  \& Privacy}, vol.~16, no.~2, pp. 52--60, 2018.

\bibitem{zennaro2020modeling}
F.~M. Zennaro and L.~Erdodi, ``Modeling penetration testing with reinforcement
  learning using capture-the-flag challenges and tabular {Q}-learning,''
  \emph{arXiv preprint arXiv:2005.12632}, 2020.

\bibitem{nist_2018}
\BIBentryALTinterwordspacing
``Framework for improving critical infrastructure cybersecurity,'' National
  Institute of Standards and Technology ({NIST}), Tech. Rep., Apr 2018.
  [Online]. Available: \url{https://doi.org/10.6028/NIST.CSWP.04162018}
\BIBentrySTDinterwordspacing

\bibitem{paxson1999bro}
V.~Paxson, ``Bro: A system for detecting network intruders in real-time,''
  \emph{Computer networks}, vol.~31, no. 23-24, pp. 2435--2463, 1999.

\bibitem{roesch1999snort}
M.~Roesch \emph{et~al.}, ``Snort: Lightweight intrusion detection for
  networks.'' in \emph{Lisa}, vol.~99, no.~1, 1999, pp. 229--238.

\bibitem{burks2012security}
D.~Burks, ``Security onion,'' \emph{Securityonion. blogspot. com}, 2012.

\bibitem{claroty}
\BIBentryALTinterwordspacing
``Claroty solution brief. tech. rep.'' 2016. [Online]. Available:
  \url{https://s3.amazonaws.com/clarotypublic/Claroty\_Solution\_Brief.pdf}
\BIBentrySTDinterwordspacing

\bibitem{tenable}
\BIBentryALTinterwordspacing
``Disrupt ot threats with tenable.ot,'' 2021. [Online]. Available:
  \url{https://www.tenable.com/products/tenable-ot}
\BIBentrySTDinterwordspacing

\bibitem{dhir2021prospective}
N.~Dhir, H.~Hoeltgebaum, N.~Adams, M.~Briers, A.~Burke, and P.~Jones,
  ``Prospective artificial intelligence approaches for active cyber defence,''
  \emph{arXiv preprint arXiv:2104.09981}, 2021.

\bibitem{gosler2013}
J.~R. Gosler and L.~Von~Thaer, ``Resilient military systems and the advanced
  cyber threat,'' \emph{Defense Science Board Technical Report}, p. 30–31,
  2013.

\bibitem{lee2016analysis}
R.~M. Lee, M.~J. Assante, and T.~Conway, ``Analysis of the cyber attack on the
  {U}krainian power grid,'' \emph{Electricity Information Sharing and Analysis
  Center (E-ISAC)}, vol. 388, 2016.

\bibitem{eaton_volz_2021}
\BIBentryALTinterwordspacing
C.~Eaton and D.~Volz, ``U.{S}. pipeline cyberattack forces closure,'' May 2021.
  [Online]. Available:
  \url{https://www.wsj.com/articles/cyberattack-forces-closure-of-largest-u-s-refined-fuel-pipeline}
\BIBentrySTDinterwordspacing

\bibitem{peiser_2021}
\BIBentryALTinterwordspacing
J.~Peiser, ``A hacker broke into a florida town's water supply and tried to
  poison it with lye, police said,'' Feb 2021. [Online]. Available:
  \url{https://www.washingtonpost.com/nation/2021/02/09/oldsmar-water-supply-hack-florida/}
\BIBentrySTDinterwordspacing

\bibitem{stouffer2011}
K.~Stouffer, J.~Falco, and K.~Scarfone, ``Guide to industrial control systems
  ({ICS}) security,'' \emph{NIST special publication}, vol. 800, no.~82, pp.
  16--16, 2011.

\bibitem{meckl2017}
S.~Meckl, G.~Tecuci, D.~Marcu, M.~Boicu, and A.~B. Zaman, ``Collaborative
  cognitive assistants for advanced persistent threat detection,'' in
  \emph{AAAI Conference on Artificial Intelligence (AAAI)}, 2017, pp. 171--178.

\bibitem{strom2016}
B.~E. Strom, A.~Applebaum, D.~P. Miller, K.~C. Nickels, A.~G. Pennington, and
  C.~B. Thomas, ``{MITRE} {ATT}\&{CK}{\textregistered}: Design and
  philosophy,'' \emph{Mitre Technical Report}, 2016.

\bibitem{williams1993}
T.~J. Williams, ``The {P}urdue enterprise reference architecture,'' in
  \emph{Information Infrastructure Systems for Manufacturing}, vol. {B-14},
  1993, pp. 43--64.

\bibitem{panda2017near}
S.~Panda and Y.~Vorobeychik, ``Near-optimal interdiction of factored {MDP}s,''
  in \emph{Conference on Uncertainty in Artificial Intelligence}, 2017.

\bibitem{elderman2017adversarial}
R.~Elderman, L.~J. Pater, A.~S. Thie, M.~M. Drugan, and M.~A. Wiering,
  ``Adversarial reinforcement learning in a cyber security simulation.'' in
  \emph{ICAART (2)}, 2017, pp. 559--566.

\bibitem{mnih2013}
V.~Mnih, K.~Kavukcuoglu, D.~Silver, A.~Graves, I.~Antonoglou, D.~Wierstra, and
  M.~A. Riedmiller, ``Playing atari with deep reinforcement learning,''
  \emph{Computing Research Repository}, 2013.

\bibitem{wei2019}
C.~Wei and T.~Ma, ``Data-dependent sample complexity of deep neural networks
  via {L}ipschitz augmentation,'' in \emph{Advances in Neural Information
  Processing Systems (NeurIPS)}, 2019, pp. 9722--9733.

\bibitem{ecoffet2021}
A.~Ecoffet, J.~Huizinga, J.~Lehman, K.~O. Stanley, and J.~Clune, ``First
  return, then explore,'' \emph{Nature}, vol. 590, no. 7847, pp. 580--586,
  2021.

\bibitem{hessel2018}
M.~Hessel, J.~Modayil, H.~van Hasselt, T.~Schaul, G.~Ostrovski, W.~Dabney,
  D.~Horgan, B.~Piot, M.~G. Azar, and D.~Silver, ``Rainbow: Combining
  improvements in deep reinforcement learning,'' in \emph{AAAI Conference on
  Artificial Intelligence (AAAI)}, 2018, pp. 3215--3222.

\bibitem{hasselt2010}
H.~van Hasselt, ``Double {Q}-learning,'' in \emph{Advances in Neural
  Information Processing Systems (NeurIPS)}, 2010, pp. 2613--2621.

\bibitem{schaul2016}
T.~Schaul, J.~Quan, I.~Antonoglou, and D.~Silver, ``Prioritized experience
  replay,'' in \emph{International Conference on Learning Representations
  (ICLR)}, 2016.

\bibitem{sutton1998}
R.~S. Sutton and A.~G. Barto, \emph{Reinforcement Learning: An Introduction},
  2nd~ed.\hskip 1em plus 0.5em minus 0.4em\relax The MIT Press, 2018.

\bibitem{ng1999}
A.~Y. Ng, D.~Harada, and S.~J. Russell, ``Policy invariance under reward
  transformations: Theory and application to reward shaping,'' in
  \emph{International Conference on Machine Learning (ICML)}, 1999, pp.
  278--287.

\bibitem{paszke2015}
A.~Paszke, S.~Gross, F.~Massa, A.~Lerer, J.~Bradbury, G.~Chanan, T.~Killeen,
  Z.~Lin, N.~Gimelshein, L.~Antiga, A.~Desmaison, A.~Kopf, E.~Yang, Z.~DeVito,
  M.~Raison, A.~Tejani, S.~Chilamkurthy, B.~Steiner, L.~Fang, J.~Bai, and
  S.~Chintala, ``Pytorch: An imperative style, high-performance deep learning
  library,'' in \emph{Advances in Neural Information Processing Systems
  (NeurIPS)}, 2019, pp. 8024--8035.

\bibitem{kullback1951}
S.~Kullback and R.~Leibler, ``{On Information and Sufficiency},'' \emph{The
  Annals of Mathematical Statistics}, vol.~22, no.~1, pp. 79 -- 86, 1951.

\bibitem{krizhevsky2012}
A.~Krizhevsky, I.~Sutskever, and G.~E. Hinton, ``Image{N}et classification with
  deep convolutional neural networks,'' in \emph{Advances in Neural Information
  Processing Systems (NeurIPS)}, 2012, pp. 1106--1114.

\bibitem{cho2014}
K.~Cho, B.~van Merrienboer, {\c{C}}.~G{\"{u}}l{\c{c}}ehre, D.~Bahdanau,
  F.~Bougares, H.~Schwenk, and Y.~Bengio, ``Learning phrase representations
  using {RNN} encoder-decoder for statistical machine translation,'' in
  \emph{Conference on Empirical Methods in Natural Language Processing
  ({EMNLP})}, 2014, pp. 1724--1734.

\bibitem{vaswani2017}
A.~Vaswani, N.~Shazeer, N.~Parmar, J.~Uszkoreit, L.~Jones, A.~N. Gomez,
  L.~Kaiser, and I.~Polosukhin, ``Attention is all you need,'' in
  \emph{Advances in Neural Information Processing Systems (NeurIPS)}, 2017, pp.
  5998--6008.

\bibitem{mern2019}
J.~Mern, D.~Sadigh, and M.~J. Kochenderfer, ``Object exchangability in
  reinforcement learning,'' in \emph{International Conference on Autonomous
  Agents and Multiagent Systems (AAMAS)}, 2019, pp. 2126--2128.

\bibitem{mern2020}
------, ``Exchangeable input representations for reinforcement learning,'' in
  \emph{American Control Conference (ACC)}, 2020, pp. 3971--3976.

\end{thebibliography}

\appendix
\section*{Network Simulator}~\label{sec: simulation}

\begin{figure}[tbh]
    \centering
    \includegraphics[width=\columnwidth]{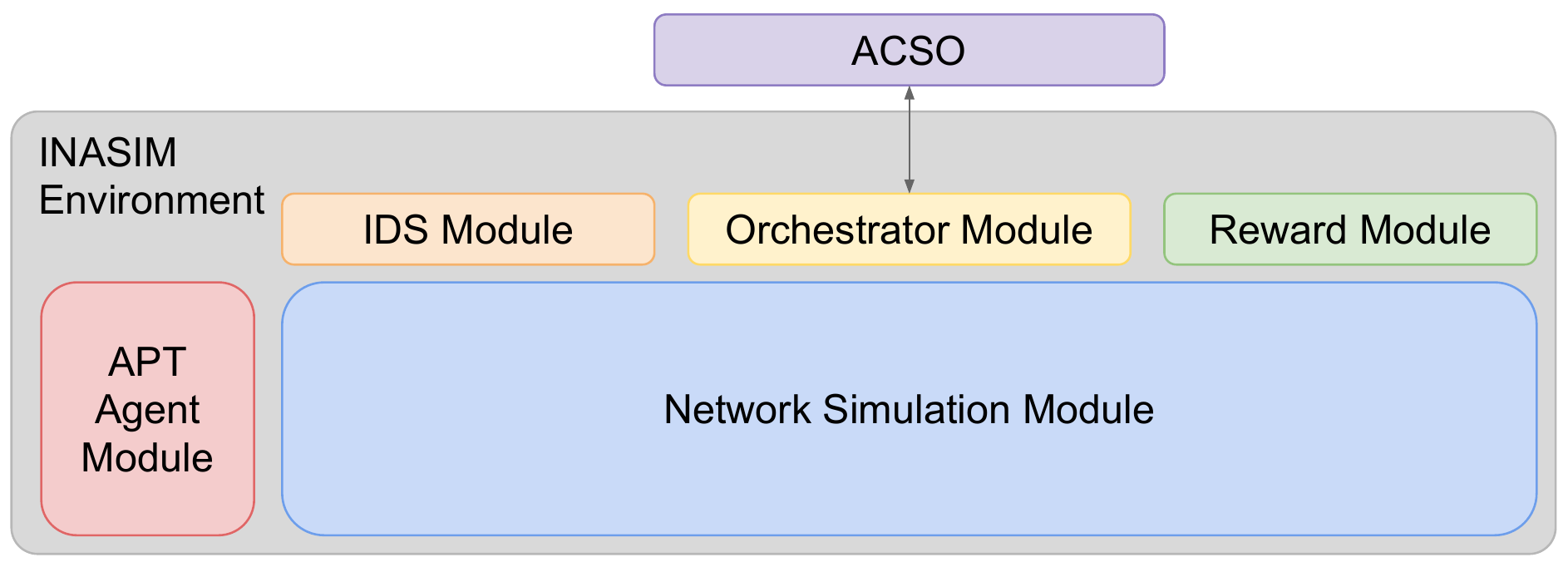}
    \caption{ICS Network Simulation Organization: The figure shows the modules composing the ICS network attack simulation. Network status and dynamics are defined by the Network Simulation Module. The APT Agent Module defines what actions the APT can take and how they affect the network dynamics. The IDS Module defines how alerts are generated from the network state. The Orchestrator Module defines what actions are available to an external ACSO agent, and the reward module allows definition of a reward function.}
    \label{fig: INASim}
\end{figure}

The ICS network attack simulator (INASIM) is organized into functional modules, as shown in~\cref{fig: INASim}. 
The code is organized around the \textit{Network Simulation} module, which defines the network structure and the dynamics of how APT and ACSO actions affect the network state. 
The \textit{APT Module} defines the actions available to the APT and provides a baseline attacker policy. 
The \textit{IDS Module} defines how alerts are generated based on the network state and agent actions. 
The \textit{Reward Module} allows a reward function to be defined for the decision making problem.

The ACSO agent interacts with the environment based on the API defined in the \textit{Orchestrator Module}. 
The orchestrator module defines which actions the ACSO can take and how these actions impact the state of the network. 
An internal API defines the interfaces between modules, allowing developers to implement new modules to simulate different attack scenarios. 

An external API allows simulations to be run with any user defined ACSO agent compatible with either the OpenAI Gym or POMDPs.jl interface. 

\subsection*{Network Simulation Module}
The network simulation module is the main element of the Gollum environment. 
The network is defined by three types of objects: nodes, networking devices, and PLCs. 
As described earlier, nodes are computing elements that APTs may compromise to spread to other nodes and to launch attacks on PLCs. 
Nodes may be workstation hosts or servers.
Nodes are connected to other nodes through networking devices. 
Each node is connected to a switch and each switch is connected to single a router. 
Each router is connected to switches or other routers through firewalls. 

Workstation nodes are homogeneous and there are three different types of server nodes.
The open platform communications (OPC) server provides direct access to scan and control the PLCs. 
The data historian server records the performance of the ICS process under control. 
In this simulation, attackers must compromise the data historian to gain knowledge of the process before executing an attack. 
The domain controller node allows network credentials to be accessed and manipulated. 
In the current simulation, however, this functionality is disabled and the domain controller is functionally equivalent to a workstation. 

All nodes are organized around switches into VLANs. 
In reality, VLANs do not require physical switches, though for simplicity each VLAN is assumed to be connected to a discrete switch, though workstation nodes can be instantly moved between VLANs.
Nodes connected to the same switch may more easily discover one another than nodes on separate VLANs. 
Each level has a dedicated router and external firewall. 
Level 1 PLCs are assumed to be connected to a level 1 switch as shown in \cref{fig:network}. 

The simulation state advances in event-driven steps. 
Each time the environment is stepped, actions generate events in a queue, and the next event is returned from the queue, along with the time at which the event occurs. 
Event steps are measured in integer units.

The state of each node on the network defines its compromise state and location.
A node's location defines which switch it is connected to in the network and is represented by an IPv4 address.
A node compromise state defines how the node has been modified by the APT, either to enable greater APT access or to prevent the ACSO from detecting or securing the intrusion. 

The full set of actions that may be taken by an APT at a given step is defined by the state of the nodes under its control and their neighbors. 
The APT has full knowledge of the compromise state of all nodes under its control. 
If a node the APT has previously scanned has been moved, it is not aware and must re-scan the node to discover it. 

\subsection*{IDS Module}
As described above, the IDS Module specifies how alerts are generated from network events and APT actions.

As described above, the IDS Module specifies how alerts are generated from network events and APT actions.
For APT actions the alert probability and level are defined by the APT action type. 
Alerts from APT actions may be generated from a single probability draw, a draw per-hour the action is taking place, or based on the message traffic generated. 
In the last case, an APT action that originates from one node and is taken on another (e.g. compromising a new node from a previously compromised node) generates message traffic across the network. 
In this case, each device that the message passes through may generate an alert.
Given a base probability of alert $p$, messages passing through a switch generate an alert with probability of $p$, messages passing through a router generate an alert with probability $2p$, and messages passing through a firewall generate an alert with probability $5p$. 

\subsection*{Orchestrator Module}
The orchestrator module defines the actions the ACSO may take and their effect on the network state. 
The ACSO may choose to take investigation and/or mitigation actions each time step, each with the effects described above (namely, investigation actions may provide alerts of compromised nodes' status while mitigation actions change the status of nodes to impede attacks). 
 
\Cref{tab: investigations} shows the investigation actions available to the defender. 
Detection probability is given for instances when malware is present. Investigations are assumed to not generate false alarms. 
Simple Scan and Human Analysis actions are taken by taking a single draw from a Bernoulli distribution with probability equal to the listed detection probability. 
If this draw is true, then an alert is generated at time equal to the current time current plus the action duration. 
Advanced scan actions continue until either an alert is created or the max duration is reached. 

\begin{table}[htb]
    \centering
    \begin{tabular}{p{0.12\linewidth} p{0.35\linewidth} p{0.1\linewidth} p{0.1\linewidth} p{0.05\linewidth}}
    \toprule
    Action & Description & Detect\newline Probability & Duration & Cost  \\
    \midrule
    Simple Scan & Simple background malware scan & 0.03/0.01 & 2 & 0.01 \\
    Advanced Scan & Disruptive malware scan & 0.05/0.02 & 8 & 0.03 \\
    Human Analysis & Task human analyst to node & 0.5/0.25 & 8 & 0.05 \\
    \bottomrule
         & 
    \end{tabular}
    \vspace{2mm}
    \caption{ACSO (Defender) Investigations. This table lists the investigation actions the ACSO may take on any node in the network. The completion time is deterministic and the alerts are generated stochastically. The detection probability is given with and without \emph{Malware Cleaned} condition.}
    \label{tab: investigations}
\end{table}

In addition to investigation actions, the ACSO can take actions to secure compromised nodes. 
These actions are described in~\cref{tab: mitigations}.
These actions will clear all compromise conditions for the targeted node unless the listed countermeasure condition is present. 
In this case, the action will have no effect. 
Additionally, the ACSO may quarantine workstation nodes. 
The Quarantine action moves the targeted node from its current VLAN to the corresponding quarantine VLAN in the same level. 
The quarantine action will return a previously quarantined node. 

\begin{table}[htb]
    \centering
    \begin{tabular}{p{0.2\linewidth} p{0.25\linewidth} p{0.2\linewidth} p{0.1\linewidth}}
        \toprule
         Action & Description & Counter- measures & Cost (host/server)\\
         \midrule
         Reboot & Power cycle node & Reboot\newline Persistence & 0.01/0.03\\
         Reset Password & Cleaned cached credentials & Credential\newline Persistence & 0.03/0.05 \\
         Re-image Node & Clean disk and reinstall OS & -- & 0.05/0.1\\
         \midrule
         Reset PLC & Reset PLC conditions & -- & 0.02 \\
         Replace PLC & Replace destroyed PLC & -- & 0.04 \\
         \bottomrule
    \end{tabular}
    \vspace{2mm}
    \caption{ACSO (Defender) Mitigations. This table lists the actions the ACSO can take to secure nodes compromised by the APT. If the countermeasures listed are not present on the targeted node, the node will be returned to nominal. The first set of actions listed can be taken on computing nodes and the second set can be taken on PLCs. The disruption cost for each action is given. For the computing node actions, the disruption costs for actions taken on workstations/servers are shown.}
    \label{tab: mitigations}
\end{table}

Action parameters such as duration and cost were solicited from a small group of cyber security researchers. 
The primary goal of the design of these parameters was to model the decision making space and trade-offs a cyber operator must make.
In real networks, effective mitigation actions are usually more costly, while less effective actions usually are less disruptive.
Our goal was to build a flexible simulation environment with parameters that capture the aspects critical to the sequential decision making problem.
These parameters could be adapted to suit specific settings as needed. 

\subsection*{APT Agent Module}
The APT Agent Module defines the actions that the APT can take and the observations the APT receives.
Any attacker policy can be used with the provided API. 
The module also provides a baseline attacker policy defined as a stochastic finite state machine (FSM).
The APT actions are given in~\cref{tab: apt actions}.
These actions are organized by which machine state uses them in the baseline policy, however, they may be called at any point by an attacker policy.
\begin{table*}[hbt]
    \centering
    \begin{tabular}{llccc}
        \toprule
        Action &  Description & Success Prob & Time Dist. & Alert Rate \\
        \midrule
        \multicolumn{5}{c}{\textit{Lateral Movement}} \\
        Scan & Scan targeted VLAN for nodes & 1.0 & (60, 0.9) & 0.01 \\
        Compromise & Gain initial control over a node & 0.9 & (60, 0.8) & 0.05 \\
        Reboot Persist & Set reboot persistence & 1.0 & (4, 0.9) & 0.05 \\
        Escalate Privilege & Gain administrator access & 1.0 & (22, 0.9) & 0.05 \\
        Credential Persist & Set credential change persistence & 1.0 & (4, 0.9) & 0.05 \\
        Cleanup & Remove malware files & 1.0 & (4, 0.9) & 0.05 \\
        \midrule
        \multicolumn{5}{c}{\textit{Vertical Movement}} \\
        Discover VLAN & Scan for occupied VLANs & 1.0 & (60, 0.9) & 0.05 \\
        Discover Server & Scan for a server on VLAN & 1.0 & (60, 0.9) & 0.01 \\
        Analyze Historian & Analyze compromised historian & 1.0 & (600, 0.9) & 0.00 \\
        \midrule
        \multicolumn{5}{c}{\textit{Attack}} \\
        Discover PLC & Scan VLAN for PLCs & 1.0 & (24, 0.875) & 0.03 \\
        Flash Firmware & Corrupt PLC firmware & 1.0 & (1, 1.0) & 0.5 \\
        Disrupt PLC & Disrupt PLC process & 1.0 & (8, 0.9) & 0.9 \\
        Destroy PLC & Destroy PLC equipment & 1.0 & (1, 1.0) & 1.0 \\
        \bottomrule
    \end{tabular}
    \vspace{2mm}
    \caption{APT Attacker Actions. This table shows the actions available to the attacker agent during the simulation. The action name and a description of the simulated effect are given. Each action attempt succeeds with the probability shown. The amount of time to complete each action is sampled from a Binomial distribution with the $n$ and $p$ parameters shown. The alert rate gives the base rate of an alert being generated by the action. For actions that generate messages on the network, this rate is multiplied by the device factor for each networking device it passes through.}
    \label{tab: apt actions}
\end{table*}
Each time step, the APT can attempt one or more of the given actions, where each action attempt succeeds with the probability shown. 
The amount of time to complete each action is sampled from a Binomial distribution taking $n$ trials each with probability $p$. 
Changes to network state caused by the action are enacted at the current time plus the duration. 
Each action attempt may generate an alert from the IDS. 
Actions occurring only on a single node generate alerts with a probability given by the alert rate.
For actions that generate messages on the network, this rate is multiplied by the device factor for each networking device it passes through.

\subsubsection*{APT Baseline Policy}
The baseline APT agent is defined by a finite state machine policy.  
Each machine state defines a stochastic sub-policy determining what actions to take in the environment and a set of transition criteria that set when to transition to a new machine state.
The FSM is parameterized by two discrete, qualitative parameters and three quantitative parameters.
The qualitative parameters are
\begin{itemize}
    \item Attack Objective: Defines whether the objective of the APT is to disrupt the ICS process or to destroy the plant equipment.
    \item Attack Vector: Defines whether the APT accesses the PLCs by compromising the OPC server or the Level 1 HMI nodes. 
\end{itemize}
The implications of the separate parameter selections are described in the main body of the work. 


\begin{figure}[tbh]
    \centering
    \includegraphics[width=\columnwidth]{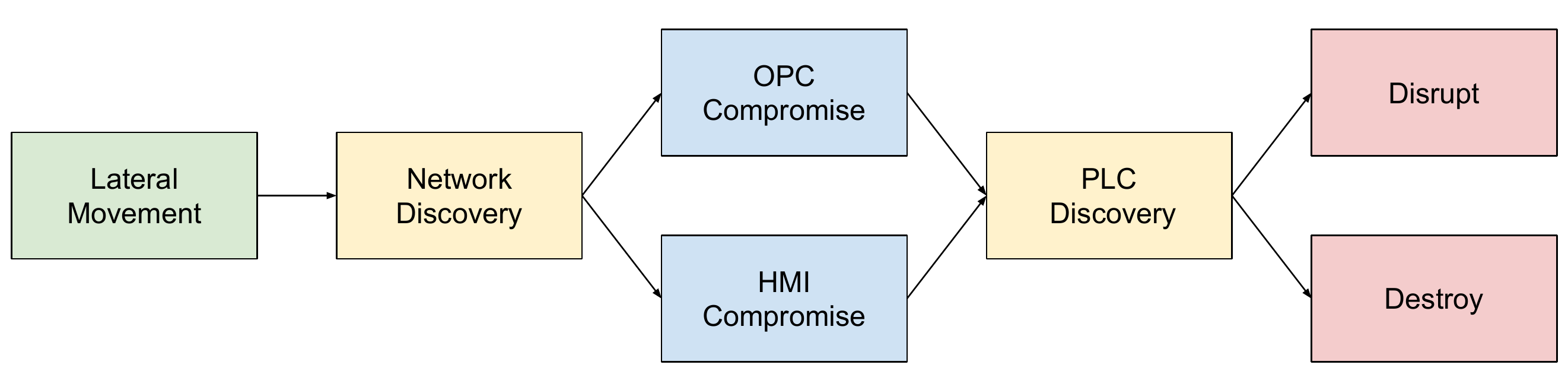}
    \caption{APT Attack Phases. This figure shows the progression of APT machine states through the course of an attack. All attacks start in lateral movement phase and progress according to the APT objective and access vector.}
    \label{fig: apt phases}
\end{figure}
There are four possible configurations of the FSM qualitative parameters which determine the progression of machine states through an attack. 
\Cref{fig: apt phases} shows the possible machine state progressions. 
Each of the APTs follows a unique policy based on which configuration it is in. 
Every APT starts in a lateral movement state and ends in a disrupt or destroy attack execution state.
The APT remains in each state until an exit criteria is met. 
If during execution, an earlier phase criteria is no longer satisfied, the policy will revert to that earlier phase before continuing. 

\section*{Neural Network Details}
The configuration of our proposed neural network is given in~\cref{tab: our config}. 
Sub-graphs for each of the three different node-types have the same dimensions. 
\begin{table}[bht]
    \centering
    \begin{tabular}{lccc}
        \toprule
        Sub-graph & No. Layers & Hidden Size & Output Size \\
        \midrule
        Attention CNN & 4 & 64 & 32 \\
        Global Attention & 2 & 128 & 128 \\
        Output MLP & 2 & 128 & -- \\        
        \bottomrule
    \end{tabular}
    \caption{Proposed Network Configuration. This table gives the parameters for each component sub-graph of the proposed attention neural network. The sub-graphs for each node type have the same dimensions.}
    \label{tab: our config}
\end{table}
A single learned vector representation is passed into a global attention graph to correspond to the output vector for the no action value. 
The MLP output heads all have tanh activation on the final layer output. 

The baseline convolutional network architecture parameters are shown in~\cref{tab: conv config}. 
\begin{table*}[htb]
    \centering
    \begin{tabular}{lccccc}
        \toprule
        Layer & Input dim. & Filter size & Stride & Output dim. & Activation \\
        \midrule
        Conv 1 & 872 & 4 & 4 & 256 & LeReLU \\
        Conv 2 & 256 & 4 & 4 & 128 & LeReLU \\
        Conv 3 & 128 & 4 & 4 & 64 & LeReLU \\
        MLP 1 & 256 & -- & -- & 256 & LeReLU \\
        Output & 256 & -- & -- & 329 & tanh \\
        \bottomrule
    \end{tabular}
    \caption{Baseline Network Configuration. The table gives the parameters for each layer of the baseline convolutional network. The convolution strides in the temporal dimension.}
    \label{tab: conv config}
\end{table*}
The baseline architecture has three 1D convolution layers, striding in the temporal dimension. 
The hidden dimensions were chosen to be as small as possible without shrinking the dimension of the latent vectors too quickly. 
The final output vector was factored into a set of vectors giving the action value for each node on the ICS network plus the value of no action. 

\section*{Training Details}
Reinforcement learning for both our proposed neural network and the baseline convolutional network was conducted using hyperparameters found via grid-search. 
The grid search was conducted on a smaller version of the problem with ten level 2 workstation nodes, three level 1 HMI nodes and thirty PLCs. 
The parameters and corresponding values searched over were 
\begin{itemize}
    \item Shaping reward weight: 0, 1, $\mathbf{\frac{1}{\gamma} = 1000}$
    \item Observation-history interval: 64, \textbf{256}
    \item Target network update frequency: \textbf{1000}, 5000, 10000
    \item $\epsilon$-greedy decay rate: 0.999, \textbf{0.9999}
\end{itemize}
where the selected parameters are shown in bold.
The performance of each parameter setting was evaluated based on the average discounted return after 500 episodes of training.
We calculated the $n$-step TD loss with $n=8$ and batches of 64.

The parameters for pretraining were selected by coordinate ascent, first testing the target margin $\delta$ with margin weighting $\lambda$ set to 0.1. 
The final values used were $\delta = 0.05$ and $\lambda = 0.1$.
The Adam optimizer was used with an initial learning rate set to $10^{-4}$ for all neural network learning.

The networks were trained using virtual machines (VM) on a cloud computing service. 
Each VM had 64 GB of RAM and 16 virtual CPUs from the Intel Sandy Bridge, Ivy Bridge, Haswell, Broadwell, and Skylake lines.
Every VM used an Nvidia Tesla T4 GPU with CUDA-enabled PyTorch v1.8.
Training 1.25 million steps of the proposed neural network took approximately 100 hours. 
All packages used were provided under open use licenses with limited restrictions, such as MIT or BSD.
Our source code is provided with an open-source MIT license. 
Details can be found in the source code license.md files. 

\section*{Playbook Policy}
\begin{figure}[htbp]
     \centering
         \includegraphics[width=\columnwidth]{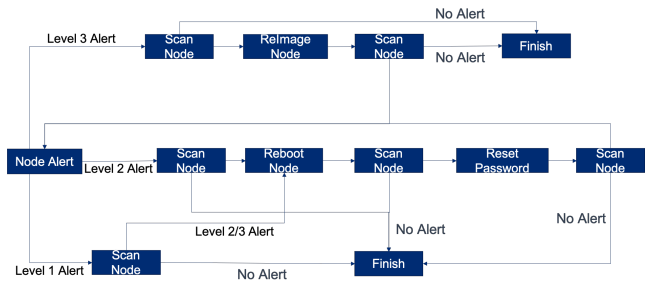}
         \caption{Playbook policy. The decision logic triggered by a node alert in the playbook policy.}
         \label{fig:playbook1}
\end{figure}

The security automation playbook baseline defines fixed responses to alerts, and is meant to be representative of current automation standards that used predefined logic to respond to triggers. 
When an alert is generated from the simulation, it triggers an automated course of action depending on the level of the severity associated with the alert.
The courses of action are comprised of scans alternating with mitigation actions based on the result of the scan, terminating when no more alerts are generated for the node.

\section*{Additional Results}

\begin{figure*}[htb]
     \centering
     \begin{subfigure}[b]{0.4\textwidth}
         \centering
         \includegraphics[width=\textwidth]{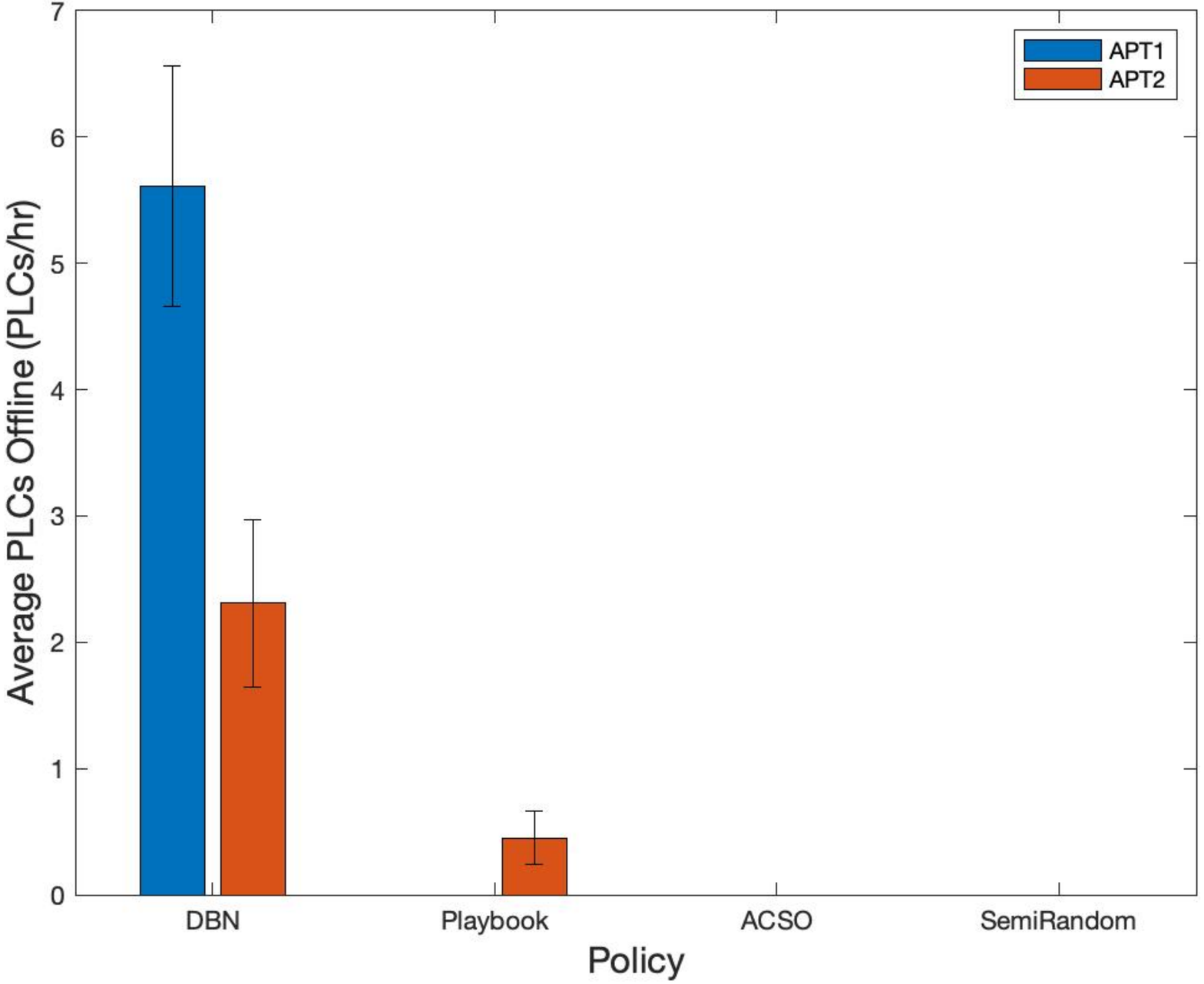}
         \caption{Final PLCs Offline}
         \label{fig:apt2_it_cost}
     \end{subfigure}
     \begin{subfigure}[b]{0.4\textwidth}
         \centering
         \includegraphics[width=\textwidth]{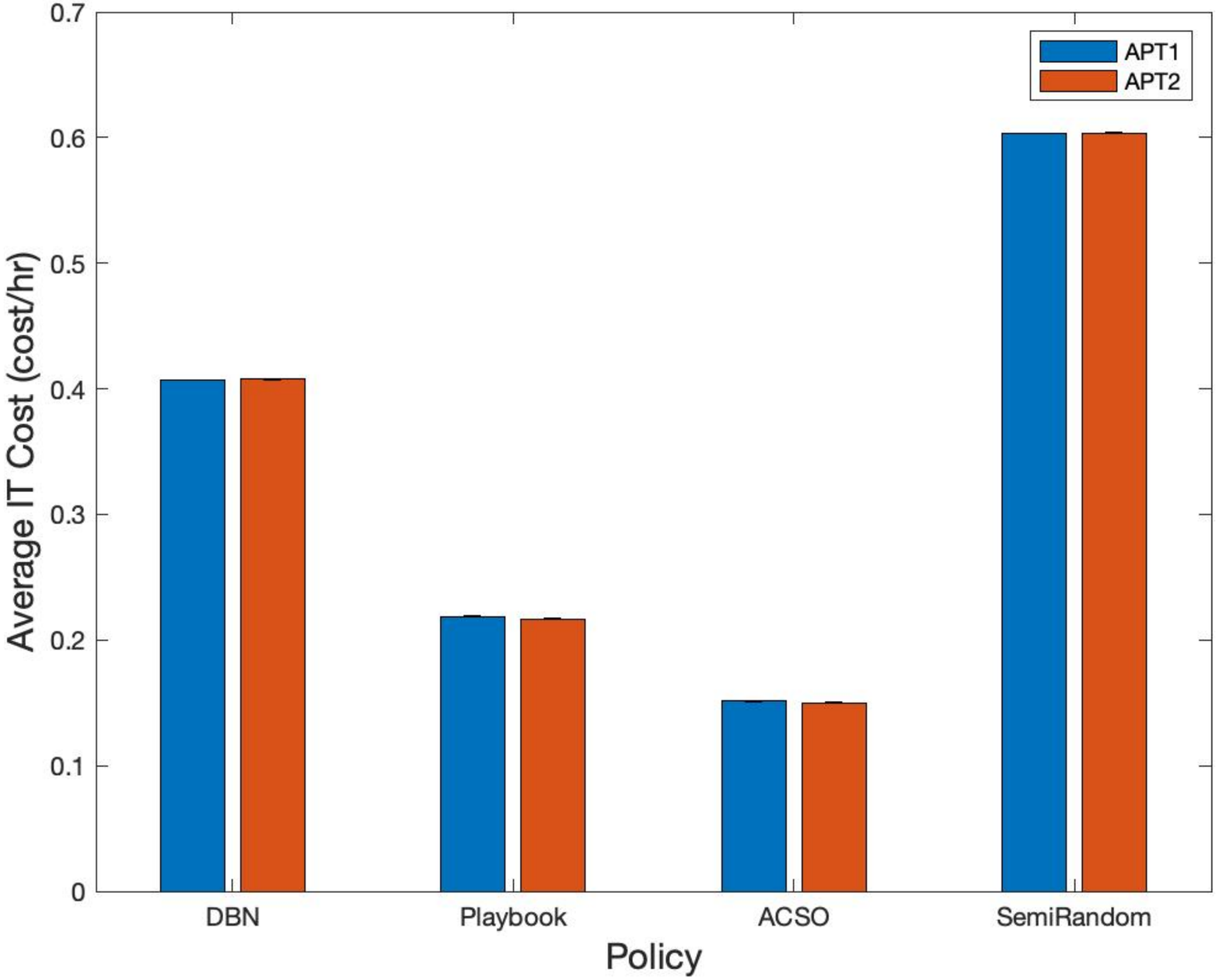}
         \caption{Average IT Cost}
         \label{fig:apt2_nodes}
     \end{subfigure}
     \begin{subfigure}[b]{0.4\textwidth}
         \centering
         \includegraphics[width=\textwidth]{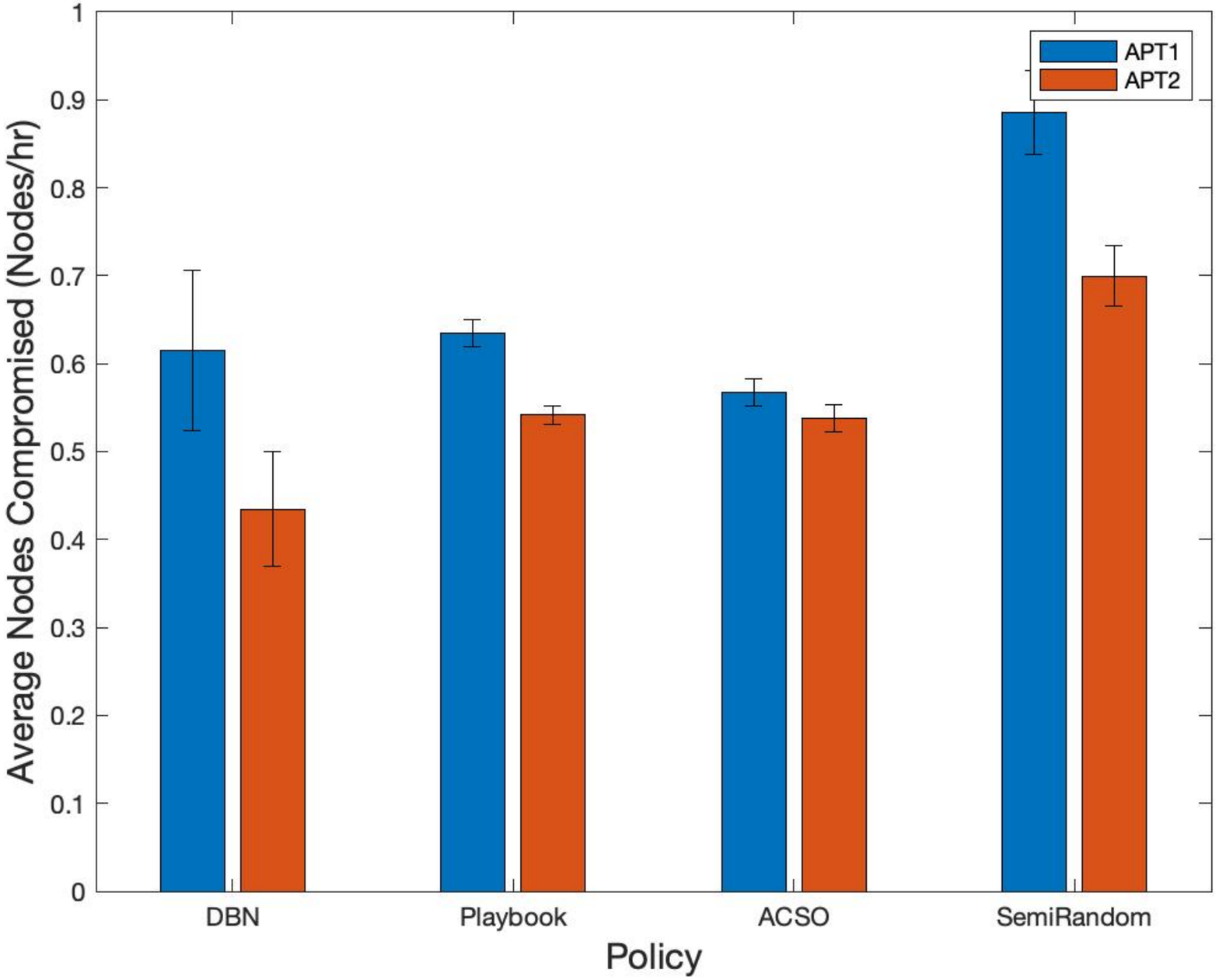}
         \caption{Average Layer 2/1 Nodes Compromised}
         \label{fig:apt2_plcs}
     \end{subfigure}
        \caption{APT Policy Experiment Results. Mean of the metric over $100$ trial episodes along with one standard error bounds for each policy in environments with different APTs. APT1 is the attacker used while training ACSO, while APT2 is a more aggressive attack policy.}
        \label{fig:apt2}
\end{figure*}

\Cref{fig:apt2} contains additional experimental results indicating the performance of the various defender policies when faced with APT1 and APT2.
As described in the results section, these indicate the robustness of the ACSO policy, particularly in comparison to the playbook policy baseline. 


\end{document}